\documentclass{aa}  

\usepackage{graphicx}
\usepackage{txfonts}
\usepackage{natbib}
\usepackage{amsmath}
\usepackage{color}
\usepackage{booktabs}
\usepackage{siunitx}
\usepackage{multirow}
\usepackage{physics}
\usepackage{xparse}
\usepackage{cuted}

\makeatletter
  \@fleqnfalse
\makeatother
\newcommand{\g}[1][1]{\gamma_{#1}}
\newcommand{\betainn}[1]{\beta_{\mathrm{in}}^{(#1)}}
\newcommand{\betaout}[1]{\beta_{\mathrm{out}}^{(#1)}}
\NewDocumentCommand{\be}{O{12} O{123}}{\ensuremath{\beta_{#1}^{(#2)}}}
\newcommand{\invisibleunderbrace}[2]{%
  \textcolor{white}{\underbrace{\textcolor{black}{#1}}_{\textcolor{black}{#2}}}%
}

\NewDocumentCommand{\twoeq}{m m o O{0pt}}{%
\noindent
\begin{minipage}{.5\linewidth}
\vspace{12pt}
\begin{align*}
\hspace{30pt}\hspace{#4} #1  
\end{align*}
\vspace{1pt}
\end{minipage}%
\begin{minipage}{.5\linewidth}
\vspace{12pt}
\refstepcounter{equation}
\begin{align*}
\hspace{-30pt}\hspace{#4} #2
\IfNoValueTF{#3}{}{\label{#3}}
\tag{\theequation}
\end{align*}
\vspace{1pt}
\end{minipage}%
}

\NewDocumentCommand{\threeeq}{m m m o O{0pt}}{%
\noindent
\begin{minipage}{.33\linewidth}
\vspace{12pt}
\begin{align*}
\hspace{30pt}\hspace{#5} #1
\end{align*}
\vspace{1pt}
\end{minipage}%
\begin{minipage}{.33\linewidth}
\vspace{12pt}
\begin{align*}
\hspace{#5} #2
\end{align*}
\vspace{1pt}
\end{minipage}%
\begin{minipage}{.33\linewidth}
\vspace{12pt}
\refstepcounter{equation}
\begin{align*}
\hspace{-30pt}\hspace{#5} #3
\IfNoValueTF{#4}{}{\label{#4}}
\tag{\theequation}
\end{align*}
\vspace{1pt}
\end{minipage}%
}

\begin{document} 
\allowdisplaybreaks

   \title{How to measure tidal dissipation in long resonant chains}
   \author{M. Cerioni\inst{1}
          \and
          C. Beaugé\inst{1}
          }

   \institute{Instituto de Astrofísica Teórica y Experimental (IATE), Observatorio de Córdoba (OAC),
              Universidad Nacional de Córdoba (UNC), Argentina,
              Córdoba, Córdoba, Laprida 854, 5000\\
              \email{matias.cerioni@unc.edu.ar}
             }

   \date{Received December 22, 2025; Accepted May 12, 2026}

 
  \abstract
   {Resonant chains are systems with three or more planets caught in a succession of two- and three-planet mean-motion resonances (2P-MMRs and 3P-MMRs). Most of the observed chains show significant amounts of separation from the nominal commensurabilities.
   These are lower energy states and therefore suggestive of a process of long-scale dissipation. The most frequently invoked plausible mechanism is active tides affecting the innermost planets, produced by the star.}
   {Simulations of tidal separation are expensive and generally impractical for extensive parameter explorations. Therefore, it is essential to have access to analytical tools that would allow us to inspect tidally separated chains, as probing these systems can give valuable insight into the physical parameters involved in dissipation.}
   {We extended an existing analytical model of the tidal separation of resonant chains with adjacent first-order 2P-MMRs that is meant to be applicable to longer $N$-planet chains. We have demonstrated how this approach can be used to constrain those parameters involved in the tidal evolution, such as the frequently unresolved $Q'$ factors.}
   {We show how this tool can be used to place meaningful bounds over the effective planetary $Q'$ value of long resonant chains, even in the realistic case where the system is poorly characterized, lacking measurements of parameters such as the stellar age or one of the planetary masses. We also show how the magnitude of separation in a resonant chain is specially sensitive to the mass of certain planets. In particular, a more massive second planet will boost tidal separation, while a more massive last planet will inhibit it.}
   {}


   \maketitle
%

\section{Introduction}

It is believed that planets form in protoplanetary disks and start migrating shortly after due to gravitational interactions with the gas medium \citep{goldreich.tremaine.1980,armitage.2020}.
In general, this process consists of an orbital decay towards the star.
During this drift, the period ratios from adjacent pairs of planets are affected.
If two converging planets encounter low-order commensurabilities, there is a chance that the planets will get captured in a two-planet mean-motion resonance (2P-MMR) \citep{ogihara.kobayashi.2013, batygin.2015, lin.liu.zheng.2025}.
Furthermore, if two adjacent pairs of planets (3 planets in total) get captured in 2P-MMRs, this will yield a three-planet MMR (3P-MMR) involving these planets, thereby forming a three-planet resonant chain.
Accordingly, any longer N-planet chain will be characterized by a succession of ($N-1$) 2P-MMRs and ($N-2$) 3P-MMRs.

\begin{figure}
\centering
\includegraphics[width=\hsize]{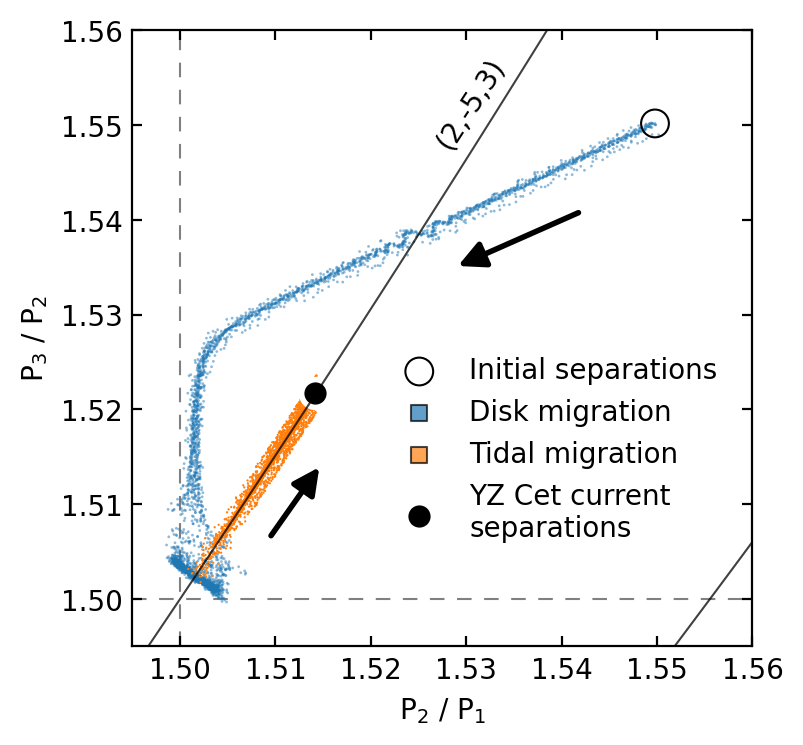}
  \caption{Simulation of the formation and tidal evolution of a three-planet resonant chain in period-ratio space. 
  Vertical and horizontal dashed lines mark 2P-MMRs, while diagonal lines mark 3P-MMRs.
  The triplet was given the initial separations marked with the empty circle.
  Blue and orange points mark two distinct phases of the evolution respectively: disk driven migration and tide driven migration.
  The black circle marks the separations of the YZ Cet, which might have been formed this way.
          }
\label{fig:chain_intro_sketch}
\end{figure}

This evolution of the dynamical structure of a planetary system can be adequately represented in period-ratio space (Fig. 1), where 2P-MMRs between the inner and outer pairs can be marked with vertical and horizontal lines, respectively, and 3P-MMRs with diagonal curves.
The blue points in Fig. \ref{fig:chain_intro_sketch} show the formation of a three-planet resonant chain via disk-driven migration, as described above.
Longer N-planet chains will produce ($N-2$) triplets in this plot.

Exoplanetary resonant chains are more readily identified by their proximity to (1) an important intersection of 2P-MMRs and (2) a zeroth-order 3P-MMR coming from it\footnote{first-order 3P-MMRs might be relevant as well \citep{cerioni.beauge.2023}, although these are, a priori much weaker.} .
Additionally, exoplanetary chains often display moderate offsets from the 2P-MMR intersection, while also remaining closely on top of the 3P-MMR.
An example of this is shown as a black dot in Fig. \ref{fig:chain_intro_sketch}, where we plot the separations of the YZ Cet system \citep{pichierri.etal.2019}.
This 2P-MMR offset stands in stark contrast with the endpoint of the disk migration process (blue curve), located much closer to the intersection.

This feature suggests a subsequent departure from the intersection in the direction of the increasing separations, towards configurations of lower orbital energy.
Such behavior is suspected to be a consequence of energy dissipation in at least one of the planets, which propagated and caused all of the planets to separate from one another.
This propagation occurs when orbits are dynamically linked by conservation of angular momentum and the 3P-MMRs (see \citealp{papaloizou.terquem.2010,batygin.morbidelli.2013b,papaloizou.2015,charalambous.etal.2018}, and also \citealp{lithwick.wu.2012, delisle.etal.2014} for the two-planet case).
These conditions pre-define the evolution paths for the separations in period-ratio space (see Section \ref{sec:pantograph}), and therefore different dissipation mechanisms only regulate the timescales.

A myriad of dissipation mechanisms have been explored in the past to explain resonant offsets, such as torques from the gas component of the protoplanetary disk (e.g. \citealp{lee.peale.2002}), disk turbulence \citep{rein.2012}, asymmetries in disk structure \citep{batygin.2015}, planet scattering with a residual disk of leftover planetesimals \citep{chatterjee.ford.2015}, or magnetospheric rebound \citep{liu.ormel.lin.2017,hansen.etal.2024}.
Perhaps the mechanism most frequently invoked in the context of resonant chains has been that of tidal friction (see works mentioned in the previous paragraph, also \citealp{pichierri.etal.2019,brasser.etal.2022,cerioni.beauge.2023, charalambous.etal.2023, charalambous.libert.2024}).

Tidal friction can be relevant after disk dissipation, when the planets stopped their disk-driven migration and weaker effects can dominate.
If the planets stopped very close to the star (orbital periods of a few days), both the star and the planets will raise tides on each other, producing friction and dissipating orbital energy as internal heat.
This effect can drive the global increase in separations and move the triplet away from the double 2P-MMR, while at the same time preserving its tight proximity to the 3P-MMR.
The orange points in Fig. \ref{fig:chain_intro_sketch} show the results of a simulation of the tidal evolution phase for the same triplet used before.

Tides active on these planets, raised by the star, can be sustained as long as the orbits are eccentric.
The resonant excitation from the chain therefore creates a promising environment for planetary tides, which likely dominate when compared to the contribution from stellar tides (see Section \ref{sec:stellar_vs_planetary} for more details).
Dissipation from planetary tides can be significantly greater if the spin axis is misaligned with the orbit normal, producing obliquity tides \citep{millholland.laughlin.2018, millholland.laughlin.2019}.
However, we note that this mechanism requires either significant mutual inclinations or very low dissipation rates (those of gas giants) to prevent rapid spin alignment (\citealp{su.lai.2022,millholland.etal.2024}).
Considering that we focus on nearly coplanar systems of low-mass planets with large offsets, we ignore obliquity tides.

The tidal evolution phase is many orders of magnitude longer than the disk-migration phase.
Resonant captures are expected to occur within the lifetime of the protoplanetary disk, with estimated ages of around $\sim 10$ Myr 
\citep{haisch.etal.2001,mamajek.2009,michel.etal.2021,pfalzner.2022,pfalzner.etal.2022,pfalzner.dincer.2024,polnitzky.etal.2026}.
In the case of the Sun, that would correspond to $\sim$0.2\% of its current lifetime; thus, the tidal evolution phase approximately covers the entirety of a system's lifetime.

As mentioned earlier in this work, many currently known resonant chains show some amount of separation from the double 2P-MMR intersection: YZ Cet \citep{stock.etal.2020}, Kepler-1530, Kepler-80 \citep{morton.etal.2016}, TOI-178 \citep{leleu.etal.2021}, K2-138 \citep{christiansen.etal.2018}, Kepler-249, and Kepler-272 \citep{rowe.etal.2014}.
Their compact scales suggest tidal friction as a likely source of dissipation (although there are low-offset exceptions which may be explained by weak tides or young stellar ages, e.g TOI-1136 or Kepler-60, see \citealp{dai.etal.2023} and \citealp{steffen.etal.2013}, respectively).
The magnitude of these separations yields valuable information on the tidal parameters, as well as other indirect factors such as stellar age and the number of planets in the chain.
In the simplest models, tides are parameterized by the so-called modified tidal quality factor, $Q'$.
This is the value we can constraint by studying the system's dynamics.

\cite{papaloizou.2015} derived an analytical model for the tidal separation of resonant chains complying with three conditions: (1) $N=3$ and (2) first-order 2P-MMRs but (3) only between adjacent planets.
This model is very useful for constraining the parameters involved in such systems, as numerical integrations of tidal evolution are generally very slow and do not allow for big grids of parameters to be adequately explored.
On the downside, the conditions for this method are quite restrictive and as of today, only one planetary system appears to be suited for its application.

In this work, we study the tidal separation of long resonant chains, presenting an extension to the model of \cite{papaloizou.2015} adapted to N-planet systems.
Although previous studies have presented frameworks for similar extensions \citep{papaloizou.2016, papaloizou.2021}, they do not solve the full system of equations to arrive at explicit solutions, as \cite{papaloizou.2015} does.
We introduce a matrix based approach to solve the full system and present the time evolution of the $n_1/n_2$ separation, useing it to show how it can place meaningful bounds on the modified tidal quality factors $Q'$ of the planets involved.
This exercise is important as $Q'$ factors are usually difficult to obtain (see \citealp{jackson.etal.2008,matsumura.etal.2008,batygin.etal.2009,hansen.2010,quinn.etal.2014,morley.etal.2017,millholland.laughlin.2018,puranam.batygin.2018,yang.wei.2022,fellay.etal.2023,mahmud.etal.2023,cerioni.beauge.2023,ferraz-mello.beauge.2025, kawai.etal.2025, shi.etal.2025}), especially for rocky planets.
We also explore the effects of high multiplicity on the tidal separation process, finding that the second and last planets play special roles in determining how much the system can open up. This study constitutes the foundation for a series of works aiming to tackle all of the above mentioned conditions.

\section{Tidal evolution}

\subsection{Period-ratio paths}\label{sec:pantograph}

While the amount of energy dissipation dictates the magnitude of the separations, the evolution paths of every orbit are strictly geometrical.
In other words, we can calculate how changes in one orbit would affect every other one. We start by considering a three-planet resonant chain composed of the resonances $n_1/n_2 \approx (p_{12}+q_{12})/p_{12}$ and $n_2/n_3 \approx (p_{23}+q_{23})/p_{23}$ between the inner and outer pair, respectively, where $n_i$ denotes the planetary mean motions. Then, the 3P-MMR and the conservation of angular momentum (up to first order in eccentricity) place the following restrictions:

\vspace{-4.15mm}

\begin{align}
    p_{12}\ n_1 - (p_{12}+q_{12}+p_{23})\ n_2 + (p_{23}+q_{23})\ n_3&=0,
    \label{eq:3pmmr} \\[5pt]
    m_1\mu^{2/3}\ n_1^{-1/3} + m_2\mu^{2/3}\ n_2^{-1/3} + m_3\mu^{2/3}\ n_3^{-1/3} &= L_0
    \label{eq:mom.ang}.
\end{align}

\noindent where $m_i$ are the masses, with sub index 0 for the star and the rest for the three planets, and $\mu=G m_0$.
Additionally, we have assumed $m_0+m_i\approx m_0$.
Differentiating both sides of each equation and approximating $\Delta \left( n_i^{-1/3} \right)\approx -1/3\, n_i^{-4/3}\, \Delta n_i$, we get two relations for the three $\Delta n_i$.
Then, we can solve for $\Delta n_2$ and $\Delta n_3$ via

\begin{align}
    \Delta n_2 &= \Delta n_1 \left(\frac{a - c\left(\frac{m_1}{m_2}\alpha_{12}^2\right)\left(\frac{m_2}{m_3}\alpha_{23}^2\right)}{b+c\left(\frac{m_2}{m_3}\alpha_{23}^2\right)}\right),
    \label{eq:delta_n2} \\
    \Delta n_3 &= -\Delta n_1 \left(\frac{a\left(\frac{m_2}{m_3}\alpha_{23}^2\right) + b\left(\frac{m_1}{m_2}\alpha_{12}^2\right)\left(\frac{m_2}{m_3}\alpha_{23}^2\right)}{b+c\left(\frac{m_2}{m_3}\alpha_{23}^2\right)}\right)
    \label{eq:delta_n3},
\end{align}

\noindent where $\alpha_{ij}=a_i/a_j$ and $a$, $b$ and $c$ are the 3P-MMR indexes $p_{12}$, $(p_{12}+q_{12}+p_{23})$ and $(p_{23}+q_{23})$, respectively.
Here we can see how changes in one orbit propagate to every other orbit\footnote{We draw a similarity to a pantograph mechanism, where moving any two nodes closer together or farther apart will simultaneously move every other pair of nodes accordingly.}.
One possibility for this to occur would be if the first planet suffered some amount of energy dissipation that caused it to fall inward.

We can see from Equation \eqref{eq:delta_n3} that the last planet would suffer a counter intuitive outward migration in reaction, as noted by the minus sign.
The migration of the middle planet, described by Equation \eqref{eq:delta_n2}, is less trivial, as it can go either way.
In the case of similar-mass planets, the negative term is proportional to $\alpha^4$, so we should expect this planet to spiral inwards along the innermost one, unless the value of $m_1/m_3$ is unusually large.
The exact mass ratio at which the second planet stays in place is given by $m_3 = m_1\frac{c}{a} \,\alpha_{12}^2\,\alpha_{23}^2$.
Lower values of $m_3$ or higher values of $m_1$ would cause the second planet to migrate outwards.

In addition to the direction of migration, we can also probe the magnitudes of these orbital alterations.
If we consider a standard three-planet chain around the 3/2-3/2 intersection, with $m_0=1\ M_\odot$, and planetary masses of 1 $M_\oplus$, we get $\Delta n_2\approx0.15\,\Delta n_1$ and $\Delta n_3\approx-0.42\,\Delta n_1$. Thus, the migration of the middle planet is always less significant than the first and third because of the opposite signs in Equation \eqref{eq:delta_n2}.

In our formulation, we actually did not require the first planet to be the one driving the global separation.
These links between the mean-motions $n_i$ are purely geometrical, and describe the tracks the planets would follow upon $\Delta n$ changes in any of the three components.

The evolution paths for mean-motions ratios will follow the 3P-MMR, as implied by Equation (\ref{eq:3pmmr}).
That means that, in the space of separations (Fig. \ref{fig:chain_intro_sketch}), the triplet will stay on top of the 3P-MMR curve as separations evolve.
In the case that the first planet loses orbital energy and were to fall inwards, the inner pair separation would increase; thus, for the triplet to stay on top of the 3P-MMR, the separation from the outer pair would increase as well.
Thus, the triplet moves away from the 2P-MMR intersection, following the 3P-MMR in the direction of increasing separations.

The geometrical link via 3P-MMRs can also be extended for longer chains.
In the case of a four-planet system, we could repeat the calculations of $\Delta n_2$, $\Delta n_3$ and now also $\Delta n_4$ by adding the new 3P-MMR relation between planets 2, 3 and 4 to the system of Equations (\ref{eq:3pmmr}) and (\ref{eq:mom.ang}).
Any longer N-chain will be composed of $(N_{pl}-2)$ 3P-MMR relations, plus the equation of the conservation of angular momentum.
The resulting behavior is analogous to the three-planet case, this time for each of the $(N-2)$ triplets.

Although tidal physics depend on both the Love number $k_2$, and the tidal quality factor $Q$, tide-driven orbital changes, such as circularization or orbital decay, depend only on the combination of $3Q/2k_2 = Q'$, as seen in Equations \eqref{eq:tau_e} and \eqref{eq:tau_n}.
This is the modified tidal quality factor, and it is the quantity we can constrain by studying orbital architectures.

Among the aims of this work is estimating $Q'$ factors through the tidal separation of resonant chains.
This factor has been historically hard to measure, especially for exoplanets.
Bearing in mind large uncertainties around this parameter, a general separation of two groups has been identified for the $Q'$ values of small rocky bodies and gas giants.
The former having approximate factors of $\sim10^{2-3}$ and $\sim10^5$ for the latter\footnote{Approximately $Q'\sim$ 50, 100, 800, 2300 for the Earth, Io, Mars and the Moon \citep{murray.dermott.1999,lainey.etal.2009,jacobson.lainey.2014,williams.boggs.2015}, $Q'\sim 136000$ for Jupiter \citep{lainey.etal.2009}, and $Q'\sim 220000-290000$ for Uranus and Neptune \citep{iess.etal.2012}, although the value for Uranus may be much higher ($Q'<840000$, \citealp{banfield.murray.1992}).}.
Notwithstanding, the spread within each group is wide enough that they can get significantly close to the other group.
On the one hand, recent calculations have suggested unexpectedly high dissipation in Saturn, with a $Q'$ value of $\sim10^4$, which is much lower than the other gas giants of the Solar System \citep{lainey.etal.2012, lainey.etal.2017}, although there is still discussion around this value (see \citealp{ogilvie.2014}).
On the other hand, we know that complex planetary structures can produce wildly different $Q'$ values.
The mere presence of an ocean can produce enough dissipation to move satellites in an otherwise tidally inactive world.
This appears to be the case for both Io and the Earth, where solid Earth dissipation is put at a much lower level.
In contrast, an ocean-less planet like Venus is expected to have much lower dissipation. 
Yet friction of atmospheric tides or even an exceedingly hot interior due to no tectonic activity, could make it a rather surprisingly dissipative planet \citep{rolf.etal.2018}.

The main foci of this work comprise signs of tidal separation exhibited by exoplanetary resonant chains.
The most comparable case study in the Solar System is that of the Galilean moons around Jupiter.
For this case, satellite tides are more important than those in the central body.
In the same vein, we considered only planetary tides on resonant exoplanets raised by the star, in contrast to tidal evolution as driven by tides raised on the star by the planets. We argue this point in the following section.

\subsection{Model}

Overall, we consider tides as the mechanism driving the global separation of the resonant chains.
We adopted a simple constant time-lag model (CTL, \citealp{mignard.1979}) and neglect tides raised on the star.
The justification is given in Section \ref{sec:stellar_vs_planetary}.

We also consider the tides raised on $m_i$ by the star $m_0$. The tidal effects on the orbit will then consist of circularization and orbital decay as characterized by the timescales \citep{goldreich.soter.1966} and expressed as

\begin{align}
    \tau_{e,i} = \frac{4}{63}\frac{Q'_i}{n_i}\frac{m_i}{m_0}\left(\frac{a_i}{R_i}\right)^5\label{eq:tau_e} ,\\
    \tau_{n,i} = -\frac{4}{189}\frac{1}{e_i^2}\frac{Q'_i}{n_i}\frac{m_i}{m_0}\left(\frac{a_i}{R_i}\right)^5\label{eq:tau_n},
\end{align}

\noindent where $Q'_i$ is the modified tidal quality factor, and it depends on the tidal quality factor, $Q_i$, while the Love number, $k_{2,i}$, is given by the relation $Q'_i = (3/2)(Q_i/k_{2,i})$. The latter is a measure of the rigidity of the body, while the former is an inverse measure of the amount of internal energy dissipation due to friction.

\subsection{Stellar versus planetary tides}\label{sec:stellar_vs_planetary}

Previous studies on the tidal evolution of resonant chains have neglected stellar tides \citep{novak.etal.2003, barnes.etal.2009,terquem.papaloizou.2007, papaloizou.terquem.2010, lithwick.wu.2012, papaloizou.2015, siegel.fabrycky.2021,brasser.etal.2022, charalambous.etal.2023,cerioni.beauge.2023}. In this work, we sought to test whether this assumption is valid and how far it can be taken.
We considered a star and a planet and compare the contributions to the planetary orbital decay due to tides produced on both bodies \citep{rodriguez.etal.2011}. Subscripts 0 and 1 refer to the star and planet, respectively, via

\begin{equation}
    \langle \dot{a_1}\rangle_{\mathrm{tid}} = \langle \dot{a_1}\rangle_{\mathrm{tid,st}} + \langle \dot{a_1}\rangle_{\mathrm{tid,pl}}
\label{eq:tau_a.1},
\end{equation}

\begin{equation}
    \langle \dot{a_1}\rangle_{\mathrm{tid}} = -\frac{4}{3}\frac{n_1}{a_1^4}\left[K_{\mathrm{st}}\left(1+23e_1^2\right) + K_{\mathrm{pl}}\left(7e_1^2\right)\right]
\label{eq:tau_a.2},
\end{equation}

\noindent where

\begin{align}
K_{\mathrm{st}} &= \frac{27}{8}\frac{1}{Q'_0}\frac{m_1}{m_0} R_0^5,  &\ \ \ \ \ \ \ \ \  K_{\mathrm{pl}} &= \frac{27}{8}\frac{1}{Q'_1}\frac{m_0}{m_1} R_1^5.
\end{align}

We can now compare the ratio between both contributions,

\begin{equation}
    \frac{\langle \dot{a_1}\rangle_{\mathrm{tid,pl}}}{\langle \dot{a_1}\rangle_{\mathrm{tid,st}}} = \frac{Q_0'}{Q_1'}\left(\frac{m_0}{m_1}\right)^2\left(\frac{R_1}{R_0}\right)^5\frac{7e_1^2}{1+23e_1^2}
\label{eq:tau_a.rat}.
\end{equation}

It is evident that for circular orbits, stellar tides will be solely responsible for the orbital decay.
Thus, we must consider at what eccentricity do the planetary tides become relevant. 
Taking the left-hand side of Equation (\ref{eq:tau_a.rat}) equal to 1, we can solve for the critical eccentricity,

\begin{equation}
e_{1,c} = \left(7\frac{Q_0'}{Q_1'}\left(\frac{m_0}{m_1}\right)^2\left(\frac{R_1}{R_0}\right)^5 - 23\right)^{-1/2}
\label{eq:e_1c}.
\end{equation}

For illustrative purposes, we can consider approximate values to that of the Sun-Earth system, where $Q_0'/Q_1'\sim10^4$ , $(m_0/m_1)^2\sim10^{11}$ and $(R_1/R_0)^5\sim10^{-10}$; in this case, for the planetary tides to be relevant, the eccentricity has to be above $e_{1,c}\sim 10^{-3}$.
Eccentricities higher than this threshold would make planetary tides the dominant contribution.
The case of Jupiter-Io returns a critical eccentricity of $e_{1,c}\sim5\times10^{-3}$.
For a more representative example of the subject matter of this work, we took masses and radii from the star and the innermost planet in the K2-138  resonant chain system, and assuming a $Q'$ ratio similar to the previous cases (consistent with previous estimations, \citealp{cerioni.beauge.2023}), again we find $e_{1,c}\sim1\times10^{-3}$.

We should note that we are omitting the dissipation contribution from tides raised on the rest of the planets. 
These would boost, in a non-negligible measure, the planetary tide contribution, effectively lowering the value for the critical eccentricity even further.

Eccentricities in resonant chains are usually higher than these bounds.
Even when accounting for the large errors involved, eccentricities for the innermost components in these systems still seem consistent with at least $e_1\gtrsim 10^{-2}$.
This is the case for chains such as that in TOI-1136 \citep{beard.etal.2024}, Kepler-223 \citep{mills.etal.2016}, Kepler-60 \citep{gozdziewski.etal.2016}, YZ Cet \citep{stock.etal.2020}, and K2-138 \citep{lopez.etal.2019}.
TRAPPIST-1 b appears to have a markedly lower eccentricity when compared to the previous systems, but still around $\sim 0.003$ \citep{grimm.etal.2018, agol.etal.2021,pichierri.etal.2024}.
These elevated eccentricities are expected in resonant chains, even after many Gyrs of tidal circularization.
That is because the resonant fixed points are associated with high eccentricities when near to the nominal commensurabilities.
Coupled with tidal circularization, a nonzero equilibrium eccentricity is expected to be sustained as long as the resonant chain is held \citep{papaloizou.2015}.

Such low values for the $e_{1,c}$ threshold suggest that the contribution of planetary tides to the orbital decay would be significantly greater than that of stellar tides.
The reason is mostly due to the elevated values of the stellar $Q'$ factors, which imply much lower energy dissipation in contrast to tides that are acting on the planet.
This fact is highlighted when we examine the Earth-Moon system, where this time (and contrary to the previous examples) the central body can dissipate more energy than the satellite ($Q'_{Earth}/Q'_{Moon}\sim10^{-1}$).
Plugging the rest of the values into Equation (\ref{eq:e_1c}) we note that the argument of the square root is negative. This means that there is no eccentricity for which lunar tides would be more important than Earth tides for the orbital decay process of the Moon.

In summary, planetary tides are expected to dominate the tidal evolution, as the planets exit the disk-phase with enough eccentricity to sustain significant dissipation.
These will nonetheless decrease over time, and stellar tides may eventually become greater.
Longer chains require more energy dissipation to achieve comparable separations, so the transition to the regime of stellar tides would take even longer.
Nevertheless, one should be wary of small $m_2/m_0$ and $R_1/R_0$ ratios, as the former produces smaller values of $e_1$ (see Eq. \ref{eq:e1_eq}), and the latter boosts stellar tides (Eq. \ref{eq:tau_a.rat}).
Both can accelerate the transition to the regime of stellar tides.
In Appendix \ref{sec:appendix_stellar_vs_planetary}, we show how the relative relevance of planetary and stellar tides evolve over time for different systems.

\section{Analytical prescription for tidal separation}

Numerical simulations of the tidal evolution of an N-planet resonant chain are slow, making it impractical to explore big grids of parameters to use the tidal history of the system to constraint other parameters.
For this reason, analytical models are crucial.

\cite{papaloizou.2015} presented a model for the tidal evolution of systems complying the following conditions: (1) three-planet chain, (2) of first-order 2P-MMRs, and (3) without nonadjacent 2P-MMRs.
The last condition is surprisingly important, as it prohibits the succession of a 3/2 and a 4/3 (in any order), given that they would produce a 2/1 2P-MMR between the first and third planet, as shown by the combination $n_1/n_3 = (n_1/n_2)\,(n_2/n_3) = (3/2)\,(4/3) = 2/1$ \citep{delisle.2017}.
This is specially prohibitive given that the most relevant first-order MMRs are 2/1, 3/2, and the 4/3.

Successive works by the same author, namely \cite{papaloizou.etal.2018} and \cite{papaloizou.2021}, presented analytical frameworks on this subject, but these studies did not solve the full system of equations as \cite{papaloizou.2015} had done. 
For this reason, we extended their method to tackle the first condition such that it can be applied to resonant chains with $N>3$, and solved for the full temporal evolution of the $n_1/n_2(t)$ separation.
We followed a matrix-based approach which is well suited for the general case of any $N$-planet chain, but kept close to the original \cite{papaloizou.2015} derivation. The other two conditions ought to be tackled in future works.

\subsection{Time evolution for N-planet chains}

We present the derivation of the method in Appendix \ref{sec:appendix_method}.
In summary, we calculate the timescale $T$ after which planets 1 and 2 increase their period ratio by 1$\%$ (starting from exact commensurability, e.g, it is the time it would take for an initial $n_1/n_2\approx 3/2$ to reach $1.515$). This timescale is used for the temporal evolution of the tidal separation itself,

\begin{align}\label{eq:n1n2_solution}
    \frac{n_1}{n_2}\big(t\big) = \frac{q_{12}+1}{q_{12}}\ \Bigg(1 + \frac{1}{100}\left(\frac{t}{T}\right)^{\frac{1}{3}}\Bigg),
\end{align}

\noindent which is equivalent to the solution obtained by \cite{papaloizou.2015}, save for a constant factor in $T$ which gives this parameter the interpretation described above. 

We tested this solution with numerical simulations of the tidal evolution of systems with different numbers of planets. We first considered a nominal six-planet chain around a star of 1 $M_\odot$.
The planetary masses and radii were taken as $m=(1,\,1.75,\,2.5,\,3.25,\,4,\,4.75)\, M_\oplus$ and $R=(1,\,1.3,\,1.6,\,1.9,\,2.2,\,2.5)\, R_\oplus$, respectively.
We ran a first simulation for the disk phase, where we gave some initial separations such that the planets will migrate and form a chain of consecutive 3/2 2P-MMRs, and (2,-5,3) 3P-MMRs.
After the capture, we rescaled the system by setting $a_1$ at $0.0338\,$ AU (corresponding to that of K2-138 b), set as the starting point of the tidal process.
We assigned modified tidal quality factors of $Q' = (100,\,200,\,300,\,400,\,500,\,600)$, and set a maximum time of 5 Gyr.
We ran this system (both the disk plus tidal phases) four different times, correspondingly lowering the number of planets to $N=$ 3, 4, 5, and 6.

\begin{figure}[h]
\centering
\includegraphics[width=\linewidth]{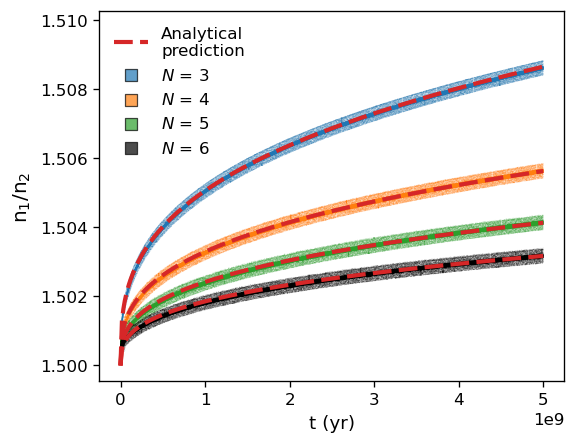}
  \caption{Tidal separation of resonant chains with different numbers of planets, $N$.
  The scatter colored points are taken from simulations (see text for details), whereas the red dashed line represents the analytical prediction as calculated with Equation \eqref{eq:n1n2_solution}.
}
     \label{fig:test_method}
\end{figure}

We present the results of the tidal simulations in Figure \ref{fig:test_method}.
We used Equation \eqref{eq:n1n2_solution} to predict the evolution of the separation and plotted it in red.
As we can see, the analytical method is surprisingly precise even after many Gyrs of evolution, and regardless of the number of planets.
We show how this method can also be applied to a currently separated chain to constrain the physical parameters allowed to achieve its separation at its current age.

\subsection{Pre-tidal offsets}\label{sec:init_offsets}
We note that we neglected initial offsets in our previous derivation.
This approximation is not trivial, as the combination of the e-damping from the gas disk and the resonant excitation of the planets will produce a nonzero equilibrium eccentricity \citep{goldreich.schlichting.2014,ramos.etal.2017, pichierri.etal.2018,choksi.chiang.2020, huang.ormel.2023} that ends up carried onto the subsequent tidal evolution. 
This eccentricity will have an associated offset, dictated by Eqs. \eqref{eq:e1_eq}-\eqref{eq:e4_eq}, where $|\Delta_{12}| \propto e_i^{-1}$.
Therefore, a small enough equilibrium eccentricity can produce an initial offset large enough to warrant specific consideration, as neglecting it would lead to an overestimation of the amount of tidal dissipation.
The simulations in Fig. \ref{fig:test_method} started with $e_1$ values between $0.006-0.01$ (increasing with the number of planets), yet the approximated analytical solutions offers a very good fit.
Notwithstanding, we should address the general validity of this approximation.

The works mentioned above considered the capture of two planets into first-order MMRs and derived the following expression for the equilibrium eccentricity of the first planet,

\begin{align}
  e_{1,eq} = \sqrt{\frac{1}{3(q_{12}+1)\mathcal{K}_1}}
  \label{eq:e1_goldreich},
\end{align}

\noindent where $\mathcal{K}_1=\tau_{n,1} /\tau_{e,1}$.
The extension to longer resonant chains of first-order MMRs (between adjacent planets only\footnote{See \citealp{delisle.2017} for the emergent complexity of first-order MMRs between nonadjacent planets which result from the consecutive combination of a 3/2 and a 4/3 resonance.}) is direct, as the equations for $\dot{e_1}$ and $\dot{n_1}$ used to derive the equilibrium eccentricity are identical, although the value of $\mathcal{K}_1$ itself will change.
Our capture simulations produced higher $e_1$ values (and lower initial offsets) for longer chains.
We can introduce the expression for $e_{1,eq}$ into Eq. \eqref{eq:e1_eq} and solve for the offset, for which we get

\begin{align}
    f_{12}(0) \sim \sqrt{3\mathcal{K}_1}\frac{m_2}{m_0}
\label{eq:init_offset},
\end{align}

\noindent where $f_{12}(t)$ is the relative mean-motion ratio separation,

\begin{align}
    f_{12}(t) = \left(\frac{n_1}{n_2}(t) - \frac{q_{12}+1}{q_{12}}\right)\left(\frac{q_{12}+1}{q_{12}}\right)^{-1}
\label{eq:f12}.
\end{align}

The magnitude of the factor $\mathcal{K}_1$ for low-mass planets (<20 $m_\oplus$) undergoing type-I migration has been estimated to lie between $\mathcal{K}_1 \sim 10^0-10^2$ from hydrodynamical simulations \citep{kley.etal.2004, cresswell.nelson.2008, bitsch.kley.2010, pichierri.etal.2023,pichierri.etal.2024}, and between $\mathcal{K}_1\sim 10^2-10^3$ from analytical estimations of a wide variety of disks \citep{ramos.etal.2017,pichierri.etal.2019,choksi.chiang.2020,fairbairn.rafikov.2025,fairbairn.etal.2025}.
Generally, $\mathcal{K}_1\propto (r/1\ \mathrm{AU})^{-2\beta}$, where $\beta$ is the flaring index.
In the disk simulations preceding those of Fig. \ref{fig:test_method}, the planets were placed in nonflared disks with a corresponding effective $\mathcal{K}_1$ value of $\sim 600$.
Initial offsets in this case were of $f_{12}(0)\approx 0.02\%$, and showed to be completely negligible, even for the six-planet chain which only reached $f_{12}(5\ \mathrm{Gyr})\approx 0.2\%$.
Nevertheless, more complicated structures such as planetary traps can significantly elevate $\mathcal{K}_1$ to $\sim 10^4$ \citep{xiang-gruess.papaloizou.2015}.
Assuming a mass ratio of $m_2/m_0 = 3\times 10^{-6}$ and taking lower and upper bounds for $\mathcal{K}$ of $10^2$ and $10^5$, we can calculate corresponding equilibrium eccentricities of $0.04$ and $0.001$, (respectively), and corresponding initial $f_{12}(0)$ offsets of $(0.005-0.2)\%$.
Although there are examples of resonant chains with little to no offset (e.g Kepler-223 or HD 110067), most chains bear current separations of $f_{12}(t)\gtrsim 1\%$.
This is the case for YZ Cet, Kepler-1530, TOI-178, Kepler-80 and K2-138 (\citealp{stock.etal.2020, holczer.etal.2016,leleu.etal.2021, shallue.vanderburg.2018,lopez.etal.2019}, respectively).
These separations are between 5-200 times larger than our largest estimates for their initial separations.

Therefore, we must consider whether these magnitudes large enough to warrant neglection of initial separations. We start by assuming a nonzero initial $f_{12}(0)$ offset in our model is equivalent to giving a head start, $t_0$, to the tidal evolution process, such that $t \rightarrow t+t_0$ (see Eq. \ref{eq:t0}).
We can approximate $t+t_0\approx t$ only if $t_0/(t+t_0)<\varepsilon$, with $\varepsilon \ll 1$.
At the same time, it can be shown that $t_0/(t+t_0) \approx (f_{12}(0)/f_{12}(t))^3$, where $f_{12}(t)$ corresponds to the current separation in a given chain.
Taking an ad hoc value of $\varepsilon=0.01$, we get the condition that $f_{12}(0)\lesssim f_{12}(t)/5$, meaning that initial separations can be neglected if these grew to be at least 5 times larger during the tidal evolution.
Our largest estimates for initial separations (corresponding to $\mathcal{K}_1 = 10^5$) produce a ratio of 5 when compared with current separations of $f_{12}(t)=1\%$, placing most currently known resonant chains within our rule even in our most restrictive case.

\begin{figure}[h]
\centering
\includegraphics[width=\linewidth]{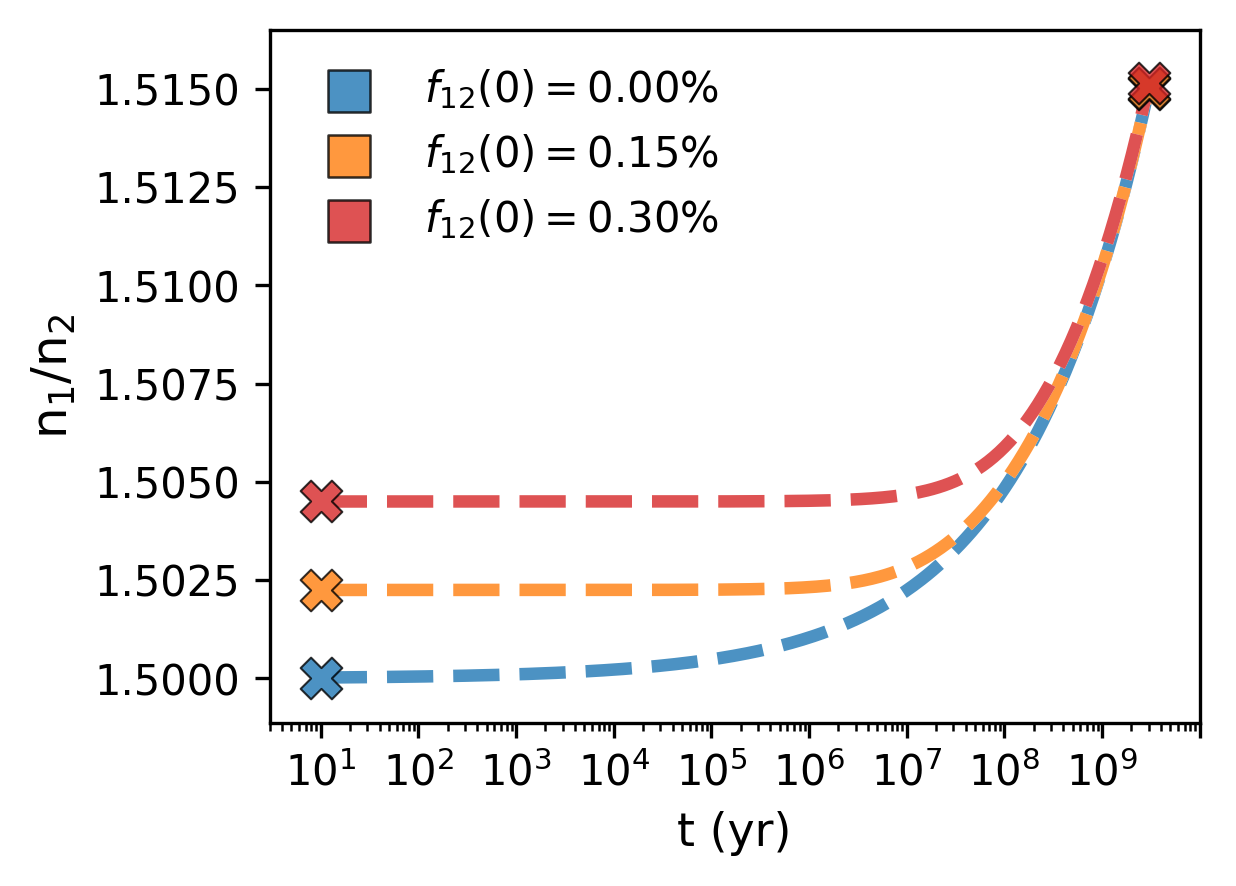}
  \caption{Tidal evolution for different values of the initial separation calculated with our analytical model.
  The system considered here comprises the five-planet system of Figure \ref{fig:test_method}, with tidal parameters chosen such that the curve with $f_{12}(0)=0.00\%$ reaches an offset of $1\%$ after 3 Gyr.}
     \label{fig:compare_init_offs}
\end{figure}

We corroborated this scenario by calculating the evolution of the separation with different values of the initial offset in Fig. \ref{fig:compare_init_offs}.
We considered a fictitious chain of five planets in consecutive 3/2 2P-MMRs and set the physical parameters such that $f_{12}(0) = 0.00\%$ and $f_{12}(3\mathrm{\ Gyr}) = 1.00\%$ (blue curve).
Additionally, we considered two more cases using the same system but assigning nonzero initial offsets of $f_{12}(0) = 0.15\%$ (orange) and $f_{12}(0) = 0.30\%$ (red) following Eq. \eqref{eq:t0}.
We can see that a modest initial offset of $0.15\%$ (1.50225) quickly converges to the curve of no initial offset after only $\sim 10$ Myr, when the $t_0/T$ term becomes insignificant.
The evolution with the larger initial offset of $0.30\%$ (1.5045) ends up converging to the blue curve as well, albeit after $\sim 100$ Myr.
Although large enough initial offsets can definitely deviate from the approximated solution, we can see that for a wide range of initial offsets, final separations (our observable) appear approximately independent of the initial ones. 
As shown above, this is expected to be the case for systems with currently large separations, which is likely to exclude young systems. 
For the remainder of this work, we neglect initial separations for simplicity, but we stress that they are still particularly important for young enough systems, which are actually expected to make up a non-negligible portion of the resonant chains \citep{hamer.schlaufman.2024, dai.etal.2024}.

\subsection{Estimating $Q'$}

For cases where the system is well characterized, we can estimate $Q'$ values by flipping Equation \eqref{eq:n1n2_solution}.
To this end, we take the constant timescale, $T$, noting that $T^{-1}$ is a linear combination of $\tau_{e,i}^{-1}$ (see Equations \ref{eq:Delta2_dDeltadt}, \ref{eq:Xij_matrix}, and \ref{eq:T}), making it in turn, a linear combination of ${Q'_i}^{-1}$.
This constitutes an approximation, given that matrix $\mathbf{M}$ (the components of which make up the coefficients of this linear combination of $\tau_{e,i}^{-1})$ also depends on the circularization timescales; however, we note that the dependence is much weaker.
We describe how we tested the validity of this approximation in Section \ref{sec:det_Qp_6pl}.

We then assume that all planets have similar planetary interiors and, as such, we can take all of the individual planetary $Q'_i$ values as equal to a single effective $Q'$.
Finally, if $T^{-1}$ is a linear combination of ${Q'_i}^{-1} = {Q'}^{-1}$, we can write $T = Q'\tilde{T}$, where $\tilde{T}=T(Q'=1)$ is be calculated from the masses, radii, and semi-major axes.

We introduce $T=Q'\tilde{T}$ into Equation \eqref{eq:n1n2_solution} and solve for $Q'$, giving

\begin{align}
    Q' = \frac{t}{\tilde{T}}\bigg(100\ f_{12}(t)\bigg)^{-3}
\label{eq:Qp_solution},
\end{align}

\noindent where $t$ now denotes the current age of the system and $f_{12}(t)$ is the currently observed relative offset, as defined in Eq. \eqref{eq:f12}.
This expression can be used to estimate $Q'$ values from separations, ages, masses, and radii.

Next we consider how this estimate relates to the $Q'_i$ values in a realistic system where these factors are not equal. We recall that $T$ is (approximately) a linear combination of $\tau_{e,i}^{-1}\propto R_i^5\, a_i^{-6.5}$ and terms with large circularization times will be negligible.
Therefore, the calculated effective $Q'$ value will result in a weighted average between the innermost planets, probably closest to $Q'_1$, although differences in radii could place it closer to $Q'_2$.
We describe how we tested the validity of this prediction in the following sections.

\section{Applicability}

The method derived in Appendix \ref{sec:appendix_method} allows us to probe the parameters involved in the tidal separation of long resonant chains; however, we note that the longest resonant chains currently known are still restrictive because they either have:
(1) second-order 2P-MMRs, for instance, K2-138 (a 3/1 between planets 5 and 6), TRAPPIST-1 (a 5/3 between planets 2 and 3) and TOI-1136 (7/5 between planets 4 and 5); or (2) nonadjacent first-order 2P-MMRs, for instance, TOI-178 (a 2/1 between planets 4 and 6), Kepler-80 (a 2/1 between planets 3 and 5 and another one between 4 and 6) or HD-110067 (a 2/1 between planets 3 and 5).
These are all six- or seven-planet systems, which are still currently out of reach.
Nonetheless, we showcase the potential applications of this tool, which can indirectly probe parameters often which are often difficult to estimate, while also showing the importance of its extension to any general $N$-planet chain.

\subsection{Determining dissipation in a long resonant chain}
\label{sec:det_Qp_6pl}

We can simulate the observation of a long resonant chain to check how precisely can this tool estimate $Q'$ values.
To this end, we considered the six-planet chain from Figure \ref{fig:test_method} at 5 Gyr, which achieved a separation of $f_{12}(5\ \mathrm{Gyr})\approx 0.21\%$ with dissipation factors of $Q'_i = (100,\,200,\,300,\,400,\,500,\,600)$.
We refer to this system with different $Q'_i$ values as \textit{DQ}. We also considered an identical system but with equal dissipation factors of $Q'_i = 100$ for every planet. Equation \eqref{eq:n1n2_solution} shows that at 5 Gyr this system will have reached a final separation of $f_{12}(5\ \mathrm{Gyr})\approx 0.24\%$.
We refer to this system as \textit{EQ}.

To simulate empirical errors on the input parameters, we assigned 5$\%$ error bands and produced lognormal distributions whose 16th, 50th, and 84th percentiles correspond to the lower bound, nominal value, and upper bound of each parameter.
We did this for the planetary masses, radii, and the stellar age. We used them to draw random values for each.
In this manner, $\sim 68\%$ of the random draws will fall inside the error bounds.
We set this level of precision for illustrative purposes, although real uncertainties are expected to be greater (specially for stellar ages).
We generated $10^5$ sets of random parameters, and calculated an effective dissipation factor $Q'$ for each, following Equation \eqref{eq:Qp_solution}.

\begin{figure}[ht!]
\centering
\includegraphics[width=\linewidth]{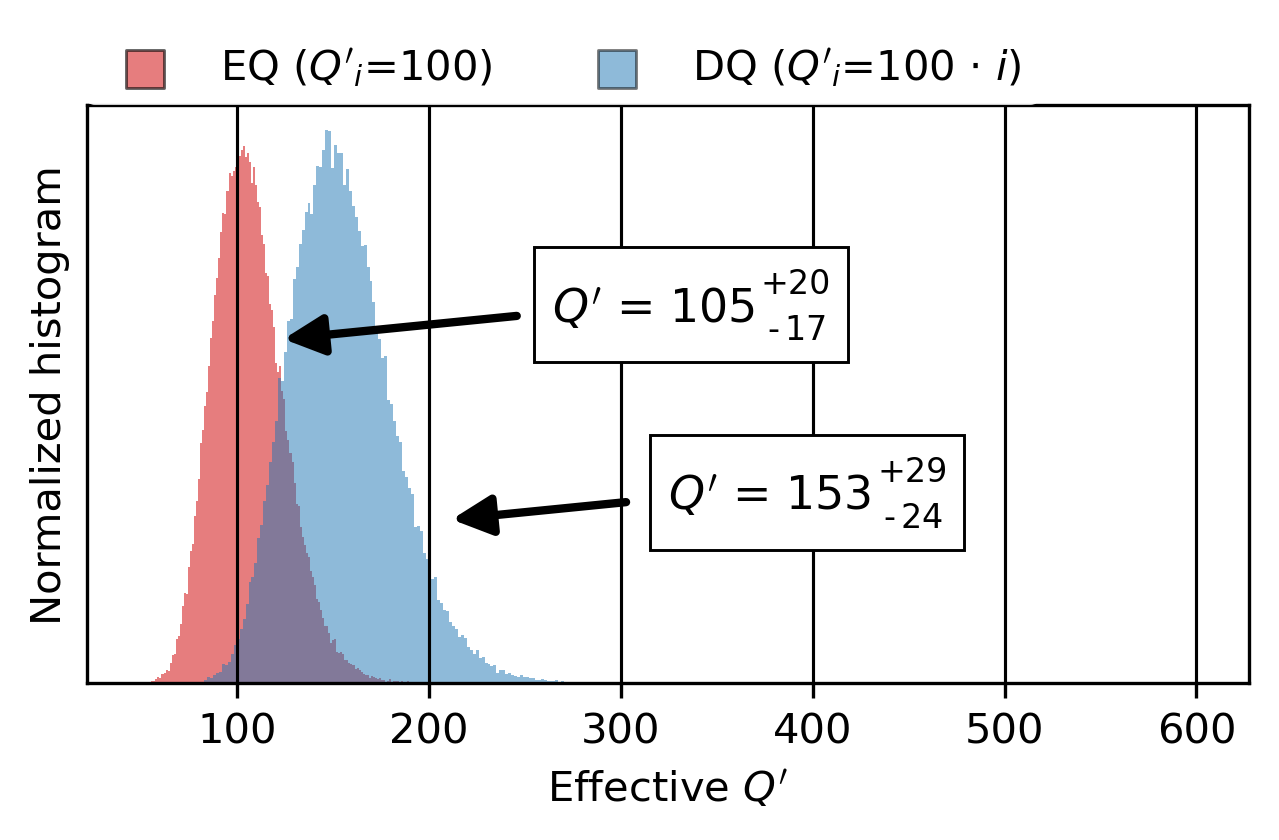}
  \caption{Resulting distribution of effective $Q'$ values for the six-planet chain shown in Figure $\ref{fig:test_method}$ (labeled DQ, blue), and an analogous system with equal dissipation factors of 100 for each of the planets (EQ, red).
  These were calculated using Equation \eqref{eq:Qp_solution} and randomly drawn values of the masses, radii, and stellar age.
  See text for details.
}
     \label{fig:det_Qp_6pl}
\end{figure}

We show the results of this experiment in Figure \ref{fig:det_Qp_6pl}, where we give the resulting distributions of effective $Q'$ values for the DQ (blue) and EQ (red) cases.
The vertical lines mark the real dissipation factors of each of the six planets in the DQ case.
We characterized the resulting distributions with a median and $1\sigma$ error band.
For the EQ case, where $Q'_i$ is equal to 100 for every planet, we recovered an effective $Q'$ value of $\sim 105^{+20}_{-17}$.
This agreement can be seen by the precise overlap of the red peak and the first vertical line.
The distribution for the DQ case returns a value of $\sim 153^{+29}_{-24}$, falling right between $Q'_1$ and $Q'_2$.
As mentioned earlier, the resulting effective $Q'$ value is expected to be representative of the innermost planets, as the relevance of each $Q'_i$ factor is weighted mainly by the semi-major axis.
For this reason, we would a priori expect this distribution to cluster closely around $Q'_1$. 
Nevertheless, the individual radii also elevate the importance of individual tides and in this system, we set $R_1<R_2$, so the peak of the distribution is closer to the middle point between $Q'_1$ and $Q'_2$.
In any case, the precision of this estimation is surprisingly informative about the dissipation of its innermost planets.

\subsection{K2-138}
\label{sec:k2-138}

\begin{figure}[h]
\centering
\includegraphics[width=\linewidth]{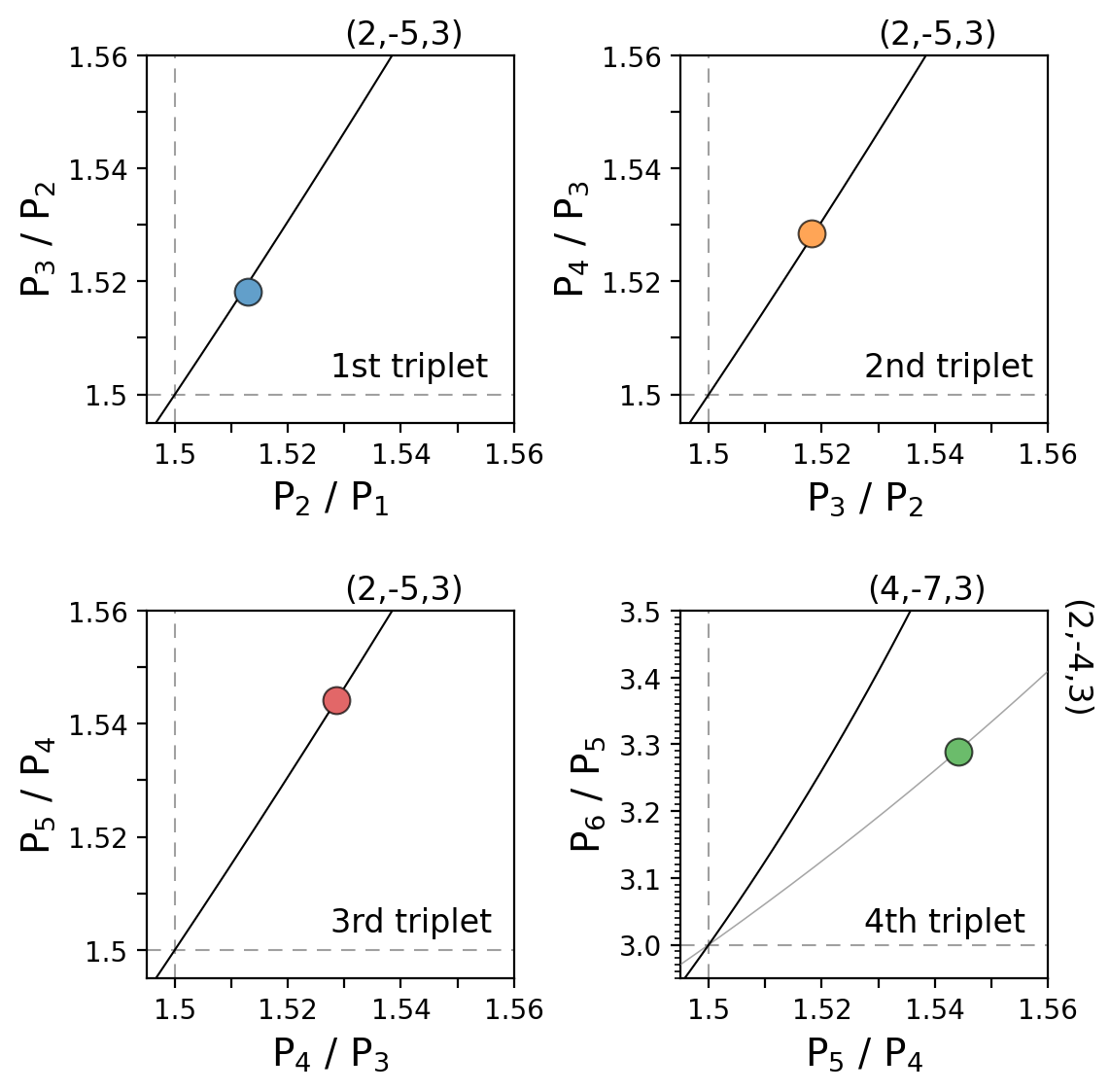}
  \caption{Resonant structure of the K2-138 system in the space of period ratios.
}
     \label{fig:k2-138_sketch}
\end{figure}

K2-138 is a K-dwarf star hosting a six-planet resonant chain \citep{cerioni.beauge.2023} involving super-Earths and sub-Neptunes.
The planets are in a succession of 2P-MMRs given by 3/2-3/2-3/2-3/2-3/1, although with sizable separations from the exact values, with successive period ratios of 1.513, 1.518, 1.529, 1.544, and 3.29, suggesting tidal separation.
We present a sketch of the resonant structure in Figure \ref{fig:k2-138_sketch}.
We select this system to show how the analytical method can be used to estimate/constraint $Q'$ values, while also serving as a benchmark to corroborate its efficacy, as there are previous estimations in the literature we can contrast with.

As we mentioned earlier, K2-138 appears to have a second-order 3/1 resonance between its last two planets.
Because of this impediment, we cannot apply our method directly and calculate $Q'$, but we can constrain it.
Omitting the outermost 2P-MMR is equivalent to ignoring the last planet and studying a a five-planet chain instead. We can predict how the addition of the sixth planet would affect the tidal evolution of the chain.
Tides active on the last planet are certainly expected to be completely irrelevant (tides decrease with $a^{-6.5}$) and, therefore, an additional outer planet would only suppress tidal separation (c.f Figure \ref{fig:test_method}).
Then, to achieve the currently observed separations at the current stellar age, a six-planet chain would have needed more dissipation (to push more planets) than its five-planet counterpart.
Therefore, a study of the five-planet case can only provide an upper bound $Q'_{max}$.

\cite{cerioni.beauge.2023} used numerical simulations to reproduce the observed separations and estimated a value of $Q'\sim 50$ for a stellar age of 2.3 Gyr \citep{lopez.etal.2019}\footnote{Although the published value is $Q'\sim 30$, we found that the tides active on the second planet were being mistakenly reduced by an error in the code and the value was corrected to $Q' \sim 50$.}.
We used the same set of parameters to compare the results and test the accuracy of the method.
A summary of the parameters adopted is displayed in the left column of Table \ref{tab:k2-138}.

\begin{table}
\setlength\tabcolsep{12pt} 
\caption{Stellar and planetary parameters for K2-138 and only its five inner planets.
}
\label{tab:k2-138}
\centering
\renewcommand{\arraystretch}{1.5} 
\begin{tabular}{lcc}
\toprule
Parameters & C23 & A22 \\
\midrule\midrule
\multicolumn{3}{l}{\textit{Stellar parameters}} \\[2.0ex]
Stellar mass ($M_\odot$)      & $0.93\pm 0.02$            & $0.891^{+0.017}_{-0.027}$ \\
Stellar radius ($R_\odot$)    & $0.86\pm 0.03$            & $0.834^{+0.011}_{-0.010}$ \\
Stellar age (Gyr)             & $2.3^{+0.44}_{-0.36}$     & $3.3^{+2.4}_{-3.2}$ \\
\midrule
\multicolumn{3}{l}{\textit{Planet b parameters}} \\[2.0ex]
\multicolumn{1}{l}{Orbital period (days)} & \multicolumn{2}{c}{\centering $2.3531$} \\
Mass ($M_\oplus$)             & $3.1^{+1.1}_{-1.1}$       & $2.80^{+0.94}_{-0.96}$ \\
Radius ($R_\oplus$)           & $1.510^{+0.110}_{-0.084}$ & $1.442^{+0.071}_{-0.063}$ \\
\midrule
\multicolumn{3}{l}{\textit{Planet c parameters}} \\[2.0ex]
\multicolumn{1}{l}{Orbital period (days)} & \multicolumn{2}{c}{\centering $3.5600$} \\
Mass ($M_\oplus$)             & $6.3^{+1.1}_{-1.2}$       & $5.95^{+1.17}_{-1.12}$ \\
Radius ($R_\oplus$)           & $2.299^{+0.120}_{-0.087}$ & $2.198^{+0.066}_{-0.054}$ \\
\midrule
\multicolumn{3}{l}{\textit{Planet d parameters}} \\[2.0ex]
\multicolumn{1}{l}{Orbital period (days)} & \multicolumn{2}{c}{\centering $5.4048$} \\
Mass ($M_\oplus$)             & $7.9^{+1.4}_{-1.3}$       & $7.20^{+1.39}_{-1.40}$ \\
Radius ($R_\oplus$)           & $2.390^{+0.104}_{-0.084}$ & $2.31^{+0.077}_{-0.068}$ \\
\midrule
\multicolumn{3}{l}{\textit{Planet e parameters}} \\[2.0ex]
\multicolumn{1}{l}{Orbital period (days)} & \multicolumn{2}{c}{\centering $8.2615$} \\
Mass ($M_\oplus$)             & 10                        & $11.28^{+2.78}_{-2.72}$ \\
Radius ($R_\oplus$)           & 3.39                      & $3.276$ \\
\midrule
\multicolumn{3}{l}{\textit{Planet f parameters}} \\[2.0ex]
\multicolumn{1}{l}{Orbital period (days)} & \multicolumn{2}{c}{\centering $12.7578$} \\
Mass ($M_\oplus$)             & 9.3                       & $2.43^{+3.05}_{-1.75}$ \\
Radius ($R_\oplus$)           & 3.013                     & $2.787$ \\
\bottomrule
\end{tabular}
\tablefoot{The columns C23 and A22 describe the two sets of parameters we used in our method, corresponding to values taken from \cite{cerioni.beauge.2023} and \cite{acuna.etal.2022}, respectively.
Orbital periods between both sets are close to identical, so we use the ones in C23 in both cases.
Values without errors are considered fixed.}
\end{table}

Next, we incorporated the uncertainties in stellar age, planetary masses and radii to our analysis by repeating the stochastic method used in Section \ref{sec:det_Qp_6pl}.
These parameters are critically important for the calculation of $Q'_{max}$, and subject to large errors as well.
We considered the error band on each parameter and produced again the corresponding lognormal distributions. We used these to generate random values of the aforementioned parameters.
For planets 4 and 5, we omitted this handling of errors for (1) their radii, given that tides are irrelevant for the outer planets and for (2) their masses, as $m_4$ and $m_5$ do not have associated errors in the source.

\begin{figure}[h]
\centering
\includegraphics[width=\linewidth]{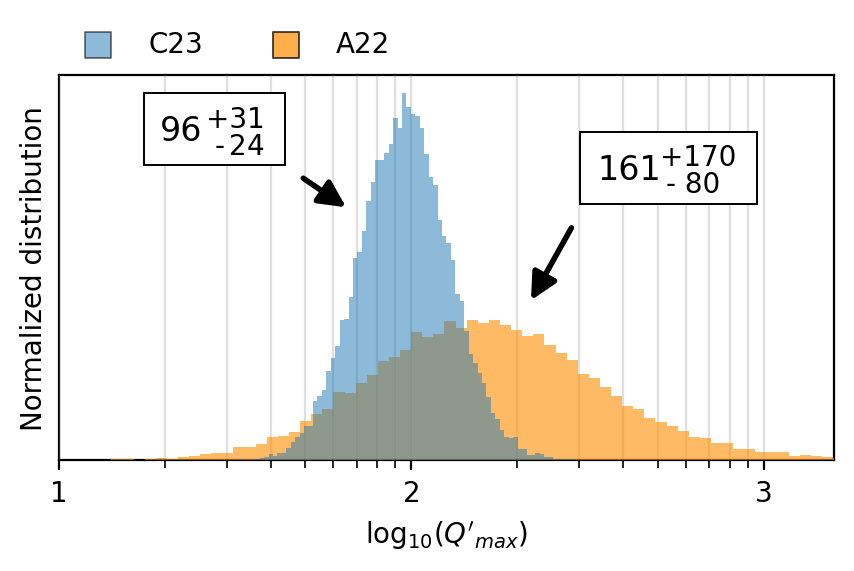}
  \caption{Resulting distributions of $Q'_{max}$ calculated from the five inner planets in the K2-138 system.
  These values are upper bounds for the real $Q'$ value associated with the complete six-planet system. These parameters were taken from two sources, C23 and A22, as noted in Table \ref{tab:k2-138}.
}
     \label{fig:k2-138_5pl_Qp}
\end{figure}

We sampled 20000 random sets of parameters and used Eq. \eqref{eq:Qp_solution} to generate a distribution of effective $Q'_{max}$. 
We present the results as the blue histogram in Figure \ref{fig:k2-138_5pl_Qp}.
The resulting distribution can be characterized by $Q'_{max}=96^{+31}_{-24}$, corresponding to the median and the 1-$\sigma$ widths on each side.
This upper bound is in agreement with the $Q'=50$ value obtained via numerical integrations in \cite{cerioni.beauge.2023}, which suggests that the method is working correctly.
Additionally, \cite{papaloizou.2021} found that a $Q'_{max} \approx 100$ would allow for tides to have significantly shaped the dynamical structure of this system.

We repeated this process using the parameters presented by \cite{acuna.etal.2022}, who reanalyzed the observations using a Bayesian scheme and arrived at different stellar and planetary parameters, which are presented in the right column of Table \ref{tab:k2-138}.
Most importantly, they obtain a stellar age of $3.3^{+2.4}_{-3.2}$ Gyr, which is significantly greater than the one we used previously ($2.3^{+0.44}_{-0.36}$ Gyr, \citealp{lopez.etal.2019}).
A longer time span to reach the observed separations means that less dissipation is required, which, in turn, indicates larger $Q'$ values.
Indeed, repeating the previous method with this set of parameters for K2-138 returns a $Q'_{max} = 161^{+170}_{-80}$, as shown in the orange histogram in Figure \ref{fig:k2-138_5pl_Qp}.

In summary, the K2-138 resonant chain seems to have been the subject of important dissipation.
This is found across different sets of the reported system parameters.
The average dissipation between the first few planets appears to be on the order of that the Earth ($Q'\sim 50$).
An extended version of the analytical method that incorporated second-order 2P-MMRs could provide a direct analytical estimation of $Q'$, instead of an upper bound.

\subsection{Attempting to characterize incomplete systems}\label{sec:role_m6}

\begin{figure*}[h]
\centering
\includegraphics[width=\textwidth]{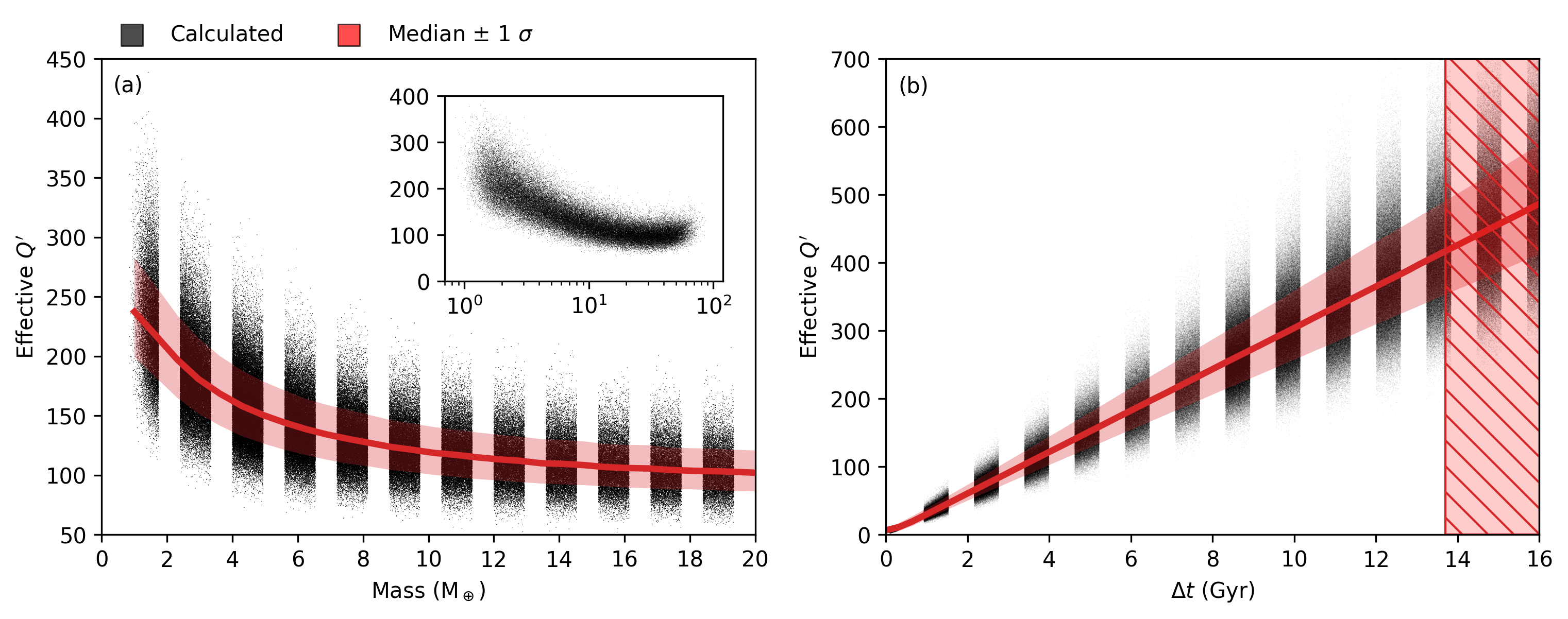}
  \caption{Effective tidal quality factors $Q'$ for a six-planet resonant chain, computed using Equation \eqref{eq:Qp_solution} and shown for two different cases of undetermined parameters: the mass of the sixth planet (left panel) and the stellar age (right panel).
  Other parameters were drawn following the stochastic method for Fig. \ref{fig:det_Qp_6pl}.
  Black dots were plotted in spaced out bins for illustrative purposes.
  The red curve marks the median of the dissipation factors at different values of the stellar age, while the surrounding red band denotes the $1\sigma$ error bar.
  For the left panel, a wider domain with a log-scale $x-$axis is shown in the subplot in the top right corner.
  For the right panel, time spans exceeding 13.7 Gyr are marked with the diagonal pattern.
  }
     \label{fig:Qp_m6_Tage}
\end{figure*}

In this section, we show how we can derive meaningful upper bounds on tidal dissipation, even upon an incomplete set of known physical parameters, such as the stellar age or one of the planetary masses.
We started with the latter.

We can examine how the individual components of the chain affect the final magnitude of tidal separation.
In particular, our tests found that the masses of the second and last planets are crucial in this regard.
A more massive second planet will boost the separation, while a less massive second planet will produce a tighter chain.
This is because the equilibrium eccentricity of the first planet is a direct function of $m_2$ (Eq. \ref{eq:e1_eq}), and this, in turn, determines the magnitude of planetary tides (Eq. \ref{eq:tau_a.2}).
The last planet, on the other hand, has to be significantly pushed outwards as the separations evolve, as shown by Eq. \eqref{eq:delta_n3}.
Therefore, the mass of the last planet should mitigate tidal separation.
In this section, we explore the dependence of the tidal evolution of a resonant chain on the mass of its last planet.
This can be informative for cases such as K2-138 where this parameter was poorly constrained \citep{lopez.etal.2019}.

We go back to the DQ system described in Section \ref{sec:det_Qp_6pl}. In this case, we followed the stochastic method used in Section \ref{sec:det_Qp_6pl}, but this time we treated $m_6$ as undetermined and sampled it from a uniform distribution bounded between 0.5 and 100 $M_\oplus$.
The radius for this planet is irrelevant for the dynamics, but it can be useful to rule out nonphysical densities, for which we give very conservative limits 0.5 and 20 $gr/cm^3$.
The rest of the parameters were drawn as described in Section \ref{sec:det_Qp_6pl}.

We plot the calculated values of effective $Q'$ in function of $m_6$ as black dots in Figure \ref{fig:Qp_m6_Tage}a.
We note that we only plotted certain bins of $m_6$ for illustrative purposes.
The red curve marks median $Q'$ values for consecutive bins of $m_6$, while the wider translucent stroke corresponds to the $1\sigma$ width.
The smaller subplot on the upper right corner shows a zoom-out to greater mass values.
As we can see, there is a decreasing relation between the mass of the last planet and the amount of dissipation required to obtain such a separation at a given time.
Our tests showed that this relation is not as narrow for the rest of the masses (except for the second planet, which also plays a special role).

The lowest values of $m_6$ are associated with the largest $Q'$ factors, which in this case give a general upper bound of $\sim 250\pm 50$, although favorable combinations of the rest of the parameters can produce factors as high as $\sim 400$.
The real mass and effective $Q'$ value corresponding to the original simulation are $(4.75,\, 153)$ and fall squarely inside the red band.
It is interesting to note that although the curve decreases for larger masses, the central curve seems to converge around $\sim 100$, not going to factors lower than 50.
Therefore, without access to information about the mass of the last planet, we can still set bounds on the effective tidal dissipation factor between $\sim 50-350$. We did not find any solutions outside of this band.


Next, we repeated this experiment but considering the stellar age $\Delta t$ as the undetermined quantity, which are usually difficult to obtain.
The details of the experiment were the same as before, but this time, the values for $\Delta t$ were taken from a uniform distribution, as this has now become the unknown quantity. With this method, we can generate $10^5$ sets of random parameters, which can be used to calculate $Q'$ values.

We present the results in Fig. \ref{fig:Qp_m6_Tage}b, in the same format as Fig. \ref{fig:Qp_m6_Tage}a.
The red diagonal pattern to the right marks those nonphysical time spans exceeding the age of the Universe.
We can see that increasing $Q'$ values naturally require longer times to reach the given separation.
The linear link between $Q'$ and $\Delta t$ was expected from Eq. \eqref{eq:tau_e}.

Without any additional knowledge of the stellar age, we only know it is bounded by the age of the Universe, which at first glance does not seem particularly informative.
However, the median curve at the left-most edge of the red area represents this upper bound and corresponds to $Q'_{\mathrm{max}}\approx 415\pm 75$ (noting that we still found some less frequent solutions reaching $600-700$).
Although greater uncertainties for the masses and radii can widen this error band, this level of constrain from such a poor bound shows that the method can be informative even for poorly characterized resonant chains.

In the previous examples, we only considered a single unknown quantity.
The determination of $Q'$ becomes degenerate when more information is missing; however, we do note that some of it is irrelevant (e.g., the radii of the last planets).
An important unknown aspect that is difficult to rule out is the presence of undetected planets on longer period orbits which are also part of the chain.
As we show in Section \ref{sec:k2-138}, the presence of additional outer planets transforms our $Q'$ estimations into upper bounds $Q'_{max}$, as the planets would have needed stronger tides to push more planets in the same timeframe.

\section{Conclusions}

In this work, we present an analytical model for the separation of $N$-planet resonant chains composed of first-order 2P-MMRs between adjacent planets, as driven by planetary tides produced by the central star.
We show how to calculate the characteristic tidal separation timescale, $T$ (see Appendix \ref{sec:appendix_method}), and used it to obtain the time evolution of the separation of the first planetary pair (Eq. \ref{eq:n1n2_solution}), linked with the rest of the pair-wise separations via the 3P-MMRs.

In general, planetary multiplicity works against tidal separation.
Additional planets at the end of the chain (where they do not contribute to the net dissipation of the system) will inhibit separation.
In particular, we show that the mass of the last planet plays a determining role in the resulting magnitude of separation.
A slight increase in mass would require a significantly greater amount of dissipation to produce the same separation.
This sensitivity is not repeated for the rest of the masses.
Although our tests show that there is an exception for the mass of the second planet, which also plays a crucial role but in the different direction.
Indeed, a more massive second planet actually boosts the amount of tidal dissipation by producing a greater equilibrium eccentricity for the first planet.

Tidal separation depends naturally on the main factors driving planetary tides, such as radii, semimajor axes, and the planetary modified tidal quality factors, $Q'_i$ (i.e., the rate of tidal dissipation inside the body).
Along with the planetary masses and the stellar age, these parameters are linked with the magnitudes of separation observed today in resonant chains.
We show how the expression for the time evolution of tidal separation can be used to place meaningful upper bounds on the amount of dissipation in very long resonant chains, even in those cases when the system is poorly characterized, lacking estimations of parameters such as the stellar age or one of the planetary masses.
Our tool can probe effective dissipation values representative of dissipation levels among the innermost planets, where tides are the most significant.
This is important because exoplanetary $Q'$ factors remain among the least frequently constrained parameters, both for individual planets and at the system-average level, especially for rocky planets (see \citealp{jackson.etal.2008,matsumura.etal.2008,batygin.etal.2009,hansen.2010,quinn.etal.2014,morley.etal.2017,millholland.laughlin.2018,puranam.batygin.2018,yang.wei.2022,fellay.etal.2023,mahmud.etal.2023,cerioni.beauge.2023,ferraz-mello.beauge.2025, kawai.etal.2025, shi.etal.2025}).

This work constitutes the foundation for a series of works aiming to extend the applicability of this method even further.
As of today, the longest known resonant chains (such as K2-138, TOI-178, Kepler-80, HD 110067 or TRAPPIST-1) are still out of reach for a direct application of this tool.
This is because of a few remaining limitations: (1) second-order 2P-MMRs, and (2) nonadjacent first-order 2P-MMRs.
Both conditions can be found in multiple known chains.
The latter one is surprisingly frequent, as it occurs when there is a succession of 4/3 and 3/2 2P-MMRs (in any order), producing a 2/1 among the first and third planet \citep{delisle.2017}.
On the other hand, indirect applications can still be considered.
By omitting the sixth planet on K2-138, we can ignore the 3/1 2P-MMR and consider the shorter chain, which is composed of successive 3/2 resonances.
Assuming similar $Q'$ factors for every planet, we obtained upper bounds of $Q'_{max}=96^{+31}_{-24}$ and $Q'_{max}=161^{+170}_{-80}$, depending on the considered set of system parameters.
The complete six-planet chain would require even more dissipation, which is why we can only obtain upper bounds. These estimations are in line with those calculated through numerical integrations for the same system \citep{cerioni.beauge.2023}.


\begin{acknowledgements}
      Most of the calculations necessary for this work were carried out with the computing facilities hosted at IATE as well as in the High Performance Computing Center of the Universidad Nacional de Córdoba (CCAD-UNC).
      This research was funded by CONICET.
\end{acknowledgements}

%
%
\bibliographystyle{aa}
\bibliography{bibliografia} 

\begin{thebibliography}{98}
\expandafter\ifx\csname natexlab\endcsname\relax\def\natexlab#1{#1}\fi

\bibitem[{{Acu{\~n}a} {et~al.}(2022){Acu{\~n}a}, {Lopez}, {Morel}, {Deleuil},
  {Mousis}, {Aguichine}, {Marcq}, \& {Santerne}}]{acuna.etal.2022}
{Acu{\~n}a}, L., {Lopez}, T.~A., {Morel}, T., {et~al.} 2022, \aap, 660, A102

\bibitem[{Agol {et~al.}(2021)Agol, Dorn, Grimm, Turbet, Ducrot, Delrez, Gillon,
  Demory, Burdanov, Barkaoui, Benkhaldoun, Bolmont, Burgasser, Carey, de~Wit,
  Fabrycky, Foreman-Mackey, Haldemann, Hernandez, Ingalls, Jehin, Langford,
  Leconte, Lederer, Luger, Malhotra, Meadows, Morris, Pozuelos, Queloz,
  Raymond, Selsis, Sestovic, Triaud, \& Grootel}]{agol.etal.2021}
Agol, E., Dorn, C., Grimm, S.~L., {et~al.} 2021, The Planetary Science Journal,
  2, 1

\bibitem[{Armitage(2020)}]{armitage.2020}
Armitage, P.~J. 2020, Astrophysics of Planet Formation, 2nd edn. (Cambridge
  University Press)

\bibitem[{{Banfield} \& {Murray}(1992)}]{banfield.murray.1992}
{Banfield}, D. \& {Murray}, N. 1992, \icarus, 99, 390

\bibitem[{{Barnes} {et~al.}(2009){Barnes}, {Jackson}, {Raymond}, {West}, \&
  {Greenberg}}]{barnes.etal.2009}
{Barnes}, R., {Jackson}, B., {Raymond}, S.~N., {West}, A.~A., \& {Greenberg},
  R. 2009, \apj, 695, 1006

\bibitem[{{Batygin}(2015)}]{batygin.2015}
{Batygin}, K. 2015, \mnras, 451, 2589

\bibitem[{{Batygin} {et~al.}(2009){Batygin}, {Bodenheimer}, \&
  {Laughlin}}]{batygin.etal.2009}
{Batygin}, K., {Bodenheimer}, P., \& {Laughlin}, G. 2009, \apjl, 704, L49

\bibitem[{{Batygin} \& {Morbidelli}(2013)}]{batygin.morbidelli.2013b}
{Batygin}, K. \& {Morbidelli}, A. 2013, \aj, 145, 1

\bibitem[{{Beard} {et~al.}(2024){Beard}, {Robertson}, {Dai}, {Holcomb},
  {Lubin}, {Akana Murphy}, {Batalha}, {Blunt}, {Crossfield}, {Dressing},
  {Fulton}, {Howard}, {Huber}, {Isaacson}, {Kane}, {Nowak}, {Petigura}, {Roy},
  {Rubenzahl}, {Weiss}, {Barrena}, {Behmard}, {Brinkman}, {Carleo}, {Chontos},
  {Dalba}, {Fetherolf}, {Giacalone}, {Hill}, {Kawauchi}, {Korth}, {Luque},
  {MacDougall}, {Mayo}, {Mo{\v{c}}nik}, {Morello}, {Murgas}, {Orell-Miquel},
  {Palle}, {Polanski}, {Rice}, {Scarsdale}, {Tyler}, \& {Van
  Zandt}}]{beard.etal.2024}
{Beard}, C., {Robertson}, P., {Dai}, F., {et~al.} 2024, \aj, 167, 70

\bibitem[{{Bitsch} \& {Kley}(2010)}]{bitsch.kley.2010}
{Bitsch}, B. \& {Kley}, W. 2010, \aap, 523, A30

\bibitem[{Brasser {et~al.}(2022)Brasser, Pichierri, Dobos, \&
  Barr}]{brasser.etal.2022}
Brasser, R., Pichierri, G., Dobos, V., \& Barr, A.~C. 2022, Monthly Notices of
  the Royal Astronomical Society, 515, 2373

\bibitem[{{Brouwer} \& {Clemence}(1961)}]{brouwer.clemence.1961}
{Brouwer}, D. \& {Clemence}, G.~M. 1961, {Methods of celestial mechanics}
  (Academic Press)

\bibitem[{{Cerioni} \& {Beaug{\'e}}(2023)}]{cerioni.beauge.2023}
{Cerioni}, M. \& {Beaug{\'e}}, C. 2023, \apj, 954, 57

\bibitem[{{Charalambous} \& {Libert}(2024)}]{charalambous.libert.2024}
{Charalambous}, C. \& {Libert}, A.-S. 2024, in IAU Symposium, Vol. 382, Complex
  Planetary Systems II: Latest Methods for an Interdisciplinary Approach, ed.
  A.~{Lemaitre} \& A.-S. {Libert}, 66--72

\bibitem[{{Charalambous} {et~al.}(2018){Charalambous}, {Mart{\'\i}},
  {Beaug{\'e}}, \& {Ramos}}]{charalambous.etal.2018}
{Charalambous}, C., {Mart{\'\i}}, J.~G., {Beaug{\'e}}, C., \& {Ramos}, X.~S.
  2018, \mnras, 477, 1414

\bibitem[{{Charalambous} {et~al.}(2023){Charalambous}, {Teyssandier}, \&
  {Libert}}]{charalambous.etal.2023}
{Charalambous}, C., {Teyssandier}, J., \& {Libert}, A.-S. 2023, \aap, 677, A160

\bibitem[{{Chatterjee} \& {Ford}(2015)}]{chatterjee.ford.2015}
{Chatterjee}, S. \& {Ford}, E.~B. 2015, \apj, 803, 33

\bibitem[{{Choksi} \& {Chiang}(2020)}]{choksi.chiang.2020}
{Choksi}, N. \& {Chiang}, E. 2020, \mnras, 495, 4192

\bibitem[{{Christiansen} {et~al.}(2018){Christiansen}, {Crossfield},
  {Barentsen}, {Lintott}, {Barclay}, {Simmons}, {Petigura}, {Schlieder},
  {Dressing}, {Vanderburg}, {Allen}, {McMaster}, {Miller}, {Veldthuis},
  {Allen}, {Wolfenbarger}, {Cox}, {Zemiro}, {Howard}, {Livingston}, {Sinukoff},
  {Catron}, {Grey}, {Kusch}, {Terentev}, {Vales}, \&
  {Kristiansen}}]{christiansen.etal.2018}
{Christiansen}, J.~L., {Crossfield}, I. J.~M., {Barentsen}, G., {et~al.} 2018,
  \aj, 155, 57

\bibitem[{{Cresswell} \& {Nelson}(2008)}]{cresswell.nelson.2008}
{Cresswell}, P. \& {Nelson}, R.~P. 2008, \aap, 482, 677

\bibitem[{{Dai} {et~al.}(2024){Dai}, {Goldberg}, {Batygin}, {van Saders},
  {Chiang}, {Choksi}, {Li}, {Petigura}, {Gilbert}, {Millholland}, {Dai},
  {Bouma}, {Weiss}, \& {Winn}}]{dai.etal.2024}
{Dai}, F., {Goldberg}, M., {Batygin}, K., {et~al.} 2024, \aj, 168, 239

\bibitem[{{Dai} {et~al.}(2023){Dai}, {Masuda}, {Beard}, {Robertson},
  {Goldberg}, {Batygin}, {Bouma}, {Lissauer}, {Knudstrup}, {Albrecht},
  {Howard}, {Knutson}, {Petigura}, {Weiss}, {Isaacson}, {Kristiansen},
  {Osborn}, {Wang}, {Wang}, {Behmard}, {Greklek-McKeon}, {Vissapragada},
  {Batalha}, {Brinkman}, {Chontos}, {Crossfield}, {Dressing}, {Fetherolf},
  {Fulton}, {Hill}, {Huber}, {Kane}, {Lubin}, {MacDougall}, {Mayo},
  {Mo{\v{c}}nik}, {Akana Murphy}, {Rubenzahl}, {Scarsdale}, {Tyler}, {Zandt},
  {Polanski}, {Schwengeler}, {Terentev}, {Benni}, {Bieryla}, {Ciardi}, {Falk},
  {Furlan}, {Girardin}, {Guerra}, {Hesse}, {Howell}, {Lillo-Box}, {Matthews},
  {Twicken}, {Villase{\~n}or}, {Latham}, {Jenkins}, {Ricker}, {Seager},
  {Vanderspek}, \& {Winn}}]{dai.etal.2023}
{Dai}, F., {Masuda}, K., {Beard}, C., {et~al.} 2023, \aj, 165, 33

\bibitem[{{Delisle}(2017)}]{delisle.2017}
{Delisle}, J.-B. 2017, \aap, 605, A96

\bibitem[{{Delisle} {et~al.}(2014){Delisle}, {Laskar}, \&
  {Correia}}]{delisle.etal.2014}
{Delisle}, J.-B., {Laskar}, J., \& {Correia}, A.~C.~M. 2014, \aap, 566, A137

\bibitem[{Fairbairn \& Dittmann(2025)}]{fairbairn.etal.2025}
Fairbairn, C.~W. \& Dittmann, A.~J. 2025, Monthly Notices of the Royal
  Astronomical Society, 543, 565

\bibitem[{{Fairbairn} \& {Rafikov}(2025)}]{fairbairn.rafikov.2025}
{Fairbairn}, C.~W. \& {Rafikov}, R.~R. 2025, \mnras, 537, 1779

\bibitem[{{Fellay} {et~al.}(2023){Fellay}, {Pezzotti}, {Buldgen},
  {Eggenberger}, \& {Bolmont}}]{fellay.etal.2023}
{Fellay}, L., {Pezzotti}, C., {Buldgen}, G., {Eggenberger}, P., \& {Bolmont},
  E. 2023, \aap, 669, A2

\bibitem[{{Ferraz-Mello} \& {Beaug{\'e}}(2025)}]{ferraz-mello.beauge.2025}
{Ferraz-Mello}, S. \& {Beaug{\'e}}, C. 2025, \aap, 697, L8

\bibitem[{{Goldreich} \& {Schlichting}(2014)}]{goldreich.schlichting.2014}
{Goldreich}, P. \& {Schlichting}, H.~E. 2014, \aj, 147, 32

\bibitem[{{Goldreich} \& {Soter}(1966)}]{goldreich.soter.1966}
{Goldreich}, P. \& {Soter}, S. 1966, \icarus, 5, 375

\bibitem[{{Goldreich} \& {Tremaine}(1980)}]{goldreich.tremaine.1980}
{Goldreich}, P. \& {Tremaine}, S. 1980, \apj, 241, 425

\bibitem[{{Go{\'z}dziewski} {et~al.}(2016){Go{\'z}dziewski}, {Migaszewski},
  {Panichi}, \& {Szuszkiewicz}}]{gozdziewski.etal.2016}
{Go{\'z}dziewski}, K., {Migaszewski}, C., {Panichi}, F., \& {Szuszkiewicz}, E.
  2016, \mnras, 455, L104

\bibitem[{{Grimm} {et~al.}(2018){Grimm}, {Demory}, {Gillon}, {Dorn}, {Agol},
  {Burdanov}, {Delrez}, {Sestovic}, {Triaud}, {Turbet}, {Bolmont}, {Caldas},
  {de Wit}, {Jehin}, {Leconte}, {Raymond}, {Van Grootel}, {Burgasser}, {Carey},
  {Fabrycky}, {Heng}, {Hernandez}, {Ingalls}, {Lederer}, {Selsis}, \&
  {Queloz}}]{grimm.etal.2018}
{Grimm}, S.~L., {Demory}, B.-O., {Gillon}, M., {et~al.} 2018, \aap, 613, A68

\bibitem[{{Haisch} {et~al.}(2001){Haisch}, {Lada}, \&
  {Lada}}]{haisch.etal.2001}
{Haisch}, Jr., K.~E., {Lada}, E.~A., \& {Lada}, C.~J. 2001, \apjl, 553, L153

\bibitem[{{Hamer} \& {Schlaufman}(2024)}]{hamer.schlaufman.2024}
{Hamer}, J.~H. \& {Schlaufman}, K.~C. 2024, \aj, 167, 55

\bibitem[{{Hansen}(2010)}]{hansen.2010}
{Hansen}, B. M.~S. 2010, \apj, 723, 285

\bibitem[{{Hansen} {et~al.}(2024){Hansen}, {Yu}, \&
  {Hasegawa}}]{hansen.etal.2024}
{Hansen}, B. M.~S., {Yu}, T.-Y., \& {Hasegawa}, Y. 2024, The Open Journal of
  Astrophysics, 7, 61

\bibitem[{{Holczer} {et~al.}(2016){Holczer}, {Mazeh}, {Nachmani},
  {Jontof-Hutter}, {Ford}, {Fabrycky}, {Ragozzine}, {Kane}, \&
  {Steffen}}]{holczer.etal.2016}
{Holczer}, T., {Mazeh}, T., {Nachmani}, G., {et~al.} 2016, \apjs, 225, 9

\bibitem[{{Huang} \& {Ormel}(2023)}]{huang.ormel.2023}
{Huang}, S. \& {Ormel}, C.~W. 2023, \mnras, 522, 828

\bibitem[{{Iess} {et~al.}(2012){Iess}, {Jacobson}, {Ducci}, {Stevenson},
  {Lunine}, {Armstrong}, {Asmar}, {Racioppa}, {Rappaport}, \&
  {Tortora}}]{iess.etal.2012}
{Iess}, L., {Jacobson}, R.~A., {Ducci}, M., {et~al.} 2012, Science, 337, 457

\bibitem[{{Jackson} {et~al.}(2008){Jackson}, {Greenberg}, \&
  {Barnes}}]{jackson.etal.2008}
{Jackson}, B., {Greenberg}, R., \& {Barnes}, R. 2008, \apj, 678, 1396

\bibitem[{{Jacobson} \& {Lainey}(2014)}]{jacobson.lainey.2014}
{Jacobson}, R.~A. \& {Lainey}, V. 2014, \planss, 102, 35

\bibitem[{{Kawai} {et~al.}(2025){Kawai}, {Fukui}, {Watanabe}, {Fukazawa}, \&
  {Narita}}]{kawai.etal.2025}
{Kawai}, Y., {Fukui}, A., {Watanabe}, N., {Fukazawa}, S., \& {Narita}, N. 2025,
  \aj, 170, 299

\bibitem[{{Kley} {et~al.}(2004){Kley}, {Peitz}, \& {Bryden}}]{kley.etal.2004}
{Kley}, W., {Peitz}, J., \& {Bryden}, G. 2004, \aap, 414, 735

\bibitem[{{Lainey} {et~al.}(2009){Lainey}, {Arlot}, {Karatekin}, \& {van
  Hoolst}}]{lainey.etal.2009}
{Lainey}, V., {Arlot}, J.-E., {Karatekin}, {\"O}., \& {van Hoolst}, T. 2009,
  \nat, 459, 957

\bibitem[{{Lainey} {et~al.}(2017){Lainey}, {Jacobson}, {Tajeddine}, {Cooper},
  {Murray}, {Robert}, {Tobie}, {Guillot}, {Mathis}, {Remus}, {Desmars},
  {Arlot}, {De Cuyper}, {Dehant}, {Pascu}, {Thuillot}, {Le Poncin-Lafitte}, \&
  {Zahn}}]{lainey.etal.2017}
{Lainey}, V., {Jacobson}, R.~A., {Tajeddine}, R., {et~al.} 2017, \icarus, 281,
  286

\bibitem[{{Lainey} {et~al.}(2012){Lainey}, {Karatekin}, {Desmars}, {Charnoz},
  {Arlot}, {Emelyanov}, {Le Poncin-Lafitte}, {Mathis}, {Remus}, {Tobie}, \&
  {Zahn}}]{lainey.etal.2012}
{Lainey}, V., {Karatekin}, {\"O}., {Desmars}, J., {et~al.} 2012, \apj, 752, 14

\bibitem[{{Lee} \& {Peale}(2002)}]{lee.peale.2002}
{Lee}, M.~H. \& {Peale}, S.~J. 2002, \apj, 567, 596

\bibitem[{{Leleu} {et~al.}(2021){Leleu}, {Alibert}, {Hara}, {Hooton}, {Wilson},
  {Robutel}, {Delisle}, {Laskar}, {Hoyer}, {Lovis}, {Bryant}, {Ducrot},
  {Cabrera}, {Delrez}, {Acton}, {Adibekyan}, {Allart}, {Allende Prieto},
  {Alonso}, {Alves}, {Anderson}, {Angerhausen}, {Anglada Escud{\'e}},
  {Asquier}, {Barrado}, {Barros}, {Baumjohann}, {Bayliss}, {Beck}, {Beck},
  {Bekkelien}, {Benz}, {Billot}, {Bonfanti}, {Bonfils}, {Bouchy}, {Bourrier},
  {Bou{\'e}}, {Brandeker}, {Broeg}, {Buder}, {Burdanov}, {Burleigh},
  {B{\'a}rczy}, {Cameron}, {Chamberlain}, {Charnoz}, {Cooke}, {Corral Van
  Damme}, {Correia}, {Cristiani}, {Damasso}, {Davies}, {Deleuil}, {Demangeon},
  {Demory}, {Di Marcantonio}, {Di Persio}, {Dumusque}, {Ehrenreich}, {Erikson},
  {Figueira}, {Fortier}, {Fossati}, {Fridlund}, {Futyan}, {Gandolfi},
  {Garc{\'\i}a Mu{\~n}oz}, {Garcia}, {Gill}, {Gillen}, {Gillon}, {Goad},
  {Gonz{\'a}lez Hern{\'a}ndez}, {Guedel}, {G{\"u}nther}, {Haldemann},
  {Henderson}, {Heng}, {Hogan}, {Isaak}, {Jehin}, {Jenkins}, {Jord{\'a}n},
  {Kiss}, {Kristiansen}, {Lam}, {Lavie}, {Lecavelier des Etangs}, {Lendl},
  {Lillo-Box}, {Lo Curto}, {Magrin}, {Martins}, {Maxted}, {McCormac}, {Mehner},
  {Micela}, {Molaro}, {Moyano}, {Murray}, {Nascimbeni}, {Nunes}, {Olofsson},
  {Osborn}, {Oshagh}, {Ottensamer}, {Pagano}, {Pall{\'e}}, {Pedersen}, {Pepe},
  {Persson}, {Peter}, {Piotto}, {Polenta}, {Pollacco}, {Poretti}, {Pozuelos},
  {Queloz}, {Ragazzoni}, {Rando}, {Ratti}, {Rauer}, {Raynard}, {Rebolo},
  {Reimers}, {Ribas}, {Santos}, {Scandariato}, {Schneider}, {Sebastian},
  {Sestovic}, {Simon}, {Smith}, {Sousa}, {Sozzetti}, {Steller}, {Su{\'a}rez
  Mascare{\~n}o}, {Szab{\'o}}, {S{\'e}gransan}, {Thomas}, {Thompson},
  {Tilbrook}, {Triaud}, {Turner}, {Udry}, {Van Grootel}, {Venus}, {Verrecchia},
  {Vines}, {Walton}, {West}, {Wheatley}, {Wolter}, \& {Zapatero
  Osorio}}]{leleu.etal.2021}
{Leleu}, A., {Alibert}, Y., {Hara}, N.~C., {et~al.} 2021, \aap, 649, A26

\bibitem[{{Lin} {et~al.}(2025){Lin}, {Liu}, \& {Zheng}}]{lin.liu.zheng.2025}
{Lin}, L., {Liu}, B., \& {Zheng}, Z. 2025, \aap, 702, A161

\bibitem[{{Lithwick} \& {Wu}(2012)}]{lithwick.wu.2012}
{Lithwick}, Y. \& {Wu}, Y. 2012, \apjl, 756, L11

\bibitem[{{Liu} {et~al.}(2017){Liu}, {Ormel}, \& {Lin}}]{liu.ormel.lin.2017}
{Liu}, B., {Ormel}, C.~W., \& {Lin}, D. N.~C. 2017, \aap, 601, A15

\bibitem[{{Lopez} {et~al.}(2019){Lopez}, {Barros}, {Santerne}, {Deleuil},
  {Adibekyan}, {Almenara}, {Armstrong}, {Brugger}, {Barrado}, {Bayliss},
  {Boisse}, {Bonomo}, {Bouchy}, {Brown}, {Carli}, {Demangeon}, {Dumusque},
  {D{\'\i}az}, {Faria}, {Figueira}, {Foxell}, {Giles}, {H{\'e}brard},
  {Hojjatpanah}, {Kirk}, {Lillo-Box}, {Lovis}, {Mousis}, {da N{\'o}brega},
  {Nielsen}, {Neal}, {Osborn}, {Pepe}, {Pollacco}, {Santos}, {Sousa}, {Udry},
  {Vigan}, \& {Wheatley}}]{lopez.etal.2019}
{Lopez}, T.~A., {Barros}, S.~C.~C., {Santerne}, A., {et~al.} 2019, \aap, 631,
  A90

\bibitem[{{Mahmud} {et~al.}(2023){Mahmud}, {Penev}, \&
  {Schussler}}]{mahmud.etal.2023}
{Mahmud}, M.~M., {Penev}, K.~M., \& {Schussler}, J.~A. 2023, \mnras, 525, 876

\bibitem[{{Mamajek}(2009)}]{mamajek.2009}
{Mamajek}, E.~E. 2009, in American Institute of Physics Conference Series, Vol.
  1158, Exoplanets and Disks: Their Formation and Diversity, ed. T.~{Usuda},
  M.~{Tamura}, \& M.~{Ishii} (AIP), 3--10

\bibitem[{{Matsumura} {et~al.}(2008){Matsumura}, {Takeda}, \&
  {Rasio}}]{matsumura.etal.2008}
{Matsumura}, S., {Takeda}, G., \& {Rasio}, F.~A. 2008, \apjl, 686, L29

\bibitem[{{Michel} {et~al.}(2021){Michel}, {van der Marel}, \&
  {Matthews}}]{michel.etal.2021}
{Michel}, A., {van der Marel}, N., \& {Matthews}, B.~C. 2021, \apj, 921, 72

\bibitem[{{Mignard}(1979)}]{mignard.1979}
{Mignard}, F. 1979, Moon and Planets, 20, 301

\bibitem[{{Millholland} \& {Laughlin}(2018)}]{millholland.laughlin.2018}
{Millholland}, S. \& {Laughlin}, G. 2018, \apjl, 869, L15

\bibitem[{{Millholland} \& {Laughlin}(2019)}]{millholland.laughlin.2019}
{Millholland}, S. \& {Laughlin}, G. 2019, Nature Astronomy, 3, 424

\bibitem[{{Millholland} {et~al.}(2024){Millholland}, {Lara}, \&
  {Toomlaid}}]{millholland.etal.2024}
{Millholland}, S.~C., {Lara}, T., \& {Toomlaid}, J. 2024, \apj, 961, 203

\bibitem[{{Mills} {et~al.}(2016){Mills}, {Fabrycky}, {Migaszewski}, {Ford},
  {Petigura}, \& {Isaacson}}]{mills.etal.2016}
{Mills}, S.~M., {Fabrycky}, D.~C., {Migaszewski}, C., {et~al.} 2016, \nat, 533,
  509

\bibitem[{{Morley} {et~al.}(2017){Morley}, {Knutson}, {Line}, {Fortney},
  {Thorngren}, {Marley}, {Teal}, \& {Lupu}}]{morley.etal.2017}
{Morley}, C.~V., {Knutson}, H., {Line}, M., {et~al.} 2017, \aj, 153, 86

\bibitem[{{Morton} {et~al.}(2016){Morton}, {Bryson}, {Coughlin}, {Rowe},
  {Ravichandran}, {Petigura}, {Haas}, \& {Batalha}}]{morton.etal.2016}
{Morton}, T.~D., {Bryson}, S.~T., {Coughlin}, J.~L., {et~al.} 2016, \apj, 822,
  86

\bibitem[{{Murray} \& {Dermott}(1999)}]{murray.dermott.1999}
{Murray}, C.~D. \& {Dermott}, S.~F. 1999, {Solar System Dynamics} (Cambridge
  University Press)

\bibitem[{{Novak} {et~al.}(2003){Novak}, {Lai}, \& {Lin}}]{novak.etal.2003}
{Novak}, G.~S., {Lai}, D., \& {Lin}, D.~N.~C. 2003, in Astronomical Society of
  the Pacific Conference Series, Vol. 294, Scientific Frontiers in Research on
  Extrasolar Planets, ed. D.~{Deming} \& S.~{Seager}, 177--180

\bibitem[{{Ogihara} \& {Kobayashi}(2013)}]{ogihara.kobayashi.2013}
{Ogihara}, M. \& {Kobayashi}, H. 2013, \apj, 775, 34

\bibitem[{{Ogilvie}(2014)}]{ogilvie.2014}
{Ogilvie}, G.~I. 2014, \araa, 52, 171

\bibitem[{{Papaloizou}(2015)}]{papaloizou.2015}
{Papaloizou}, J. C.~B. 2015, International Journal of Astrobiology, 14, 291

\bibitem[{{Papaloizou}(2016)}]{papaloizou.2016}
{Papaloizou}, J.~C.~B. 2016, Celestial Mechanics and Dynamical Astronomy, 126,
  157

\bibitem[{{Papaloizou}(2021)}]{papaloizou.2021}
{Papaloizou}, J.~C.~B. 2021, Celestial Mechanics and Dynamical Astronomy, 133,
  30

\bibitem[{{Papaloizou} {et~al.}(2018){Papaloizou}, {Szuszkiewicz}, \&
  {Terquem}}]{papaloizou.etal.2018}
{Papaloizou}, J.~C.~B., {Szuszkiewicz}, E., \& {Terquem}, C. 2018, \mnras, 476,
  5032

\bibitem[{Papaloizou \& Terquem(2010)}]{papaloizou.terquem.2010}
Papaloizou, J. C.~B. \& Terquem, C. 2010, Monthly Notices of the Royal
  Astronomical Society, 405, 573

\bibitem[{{Pfalzner}(2022)}]{pfalzner.2022}
{Pfalzner}, S. 2022, Research Notes of the American Astronomical Society, 6,
  219

\bibitem[{{Pfalzner} {et~al.}(2022){Pfalzner}, {Dehghani}, \&
  {Michel}}]{pfalzner.etal.2022}
{Pfalzner}, S., {Dehghani}, S., \& {Michel}, A. 2022, \apjl, 939, L10

\bibitem[{{Pfalzner} \& {Dincer}(2024)}]{pfalzner.dincer.2024}
{Pfalzner}, S. \& {Dincer}, F. 2024, \apj, 963, 122

\bibitem[{{Pichierri} {et~al.}(2019){Pichierri}, {Batygin}, \&
  {Morbidelli}}]{pichierri.etal.2019}
{Pichierri}, G., {Batygin}, K., \& {Morbidelli}, A. 2019, \aap, 625, A7

\bibitem[{{Pichierri} {et~al.}(2023){Pichierri}, {Bitsch}, \&
  {Lega}}]{pichierri.etal.2023}
{Pichierri}, G., {Bitsch}, B., \& {Lega}, E. 2023, \aap, 670, A148

\bibitem[{{Pichierri} {et~al.}(2024){Pichierri}, {Morbidelli}, {Batygin}, \&
  {Brasser}}]{pichierri.etal.2024}
{Pichierri}, G., {Morbidelli}, A., {Batygin}, K., \& {Brasser}, R. 2024, Nature
  Astronomy, 8, 1408

\bibitem[{{Pichierri} {et~al.}(2018){Pichierri}, {Morbidelli}, \&
  {Crida}}]{pichierri.etal.2018}
{Pichierri}, G., {Morbidelli}, A., \& {Crida}, A. 2018, Celestial Mechanics and
  Dynamical Astronomy, 130, 54

\bibitem[{{Polnitzky} {et~al.}(2026){Polnitzky}, {Ratzenb{\"o}ck},
  {Gro{\ss}schedl}, \& {Alves}}]{polnitzky.etal.2026}
{Polnitzky}, F.~A., {Ratzenb{\"o}ck}, S., {Gro{\ss}schedl}, J., \& {Alves}, J.
  2026, \aap, 707, A216

\bibitem[{{Puranam} \& {Batygin}(2018)}]{puranam.batygin.2018}
{Puranam}, A. \& {Batygin}, K. 2018, \aj, 155, 157

\bibitem[{{Quinn} {et~al.}(2014){Quinn}, {White}, {Latham}, {Buchhave},
  {Torres}, {Stefanik}, {Berlind}, {Bieryla}, {Calkins}, {Esquerdo},
  {F{\H{u}}r{\'e}sz}, {Geary}, \& {Szentgyorgyi}}]{quinn.etal.2014}
{Quinn}, S.~N., {White}, R.~J., {Latham}, D.~W., {et~al.} 2014, \apj, 787, 27

\bibitem[{{Ramos} {et~al.}(2017){Ramos}, {Charalambous},
  {Ben{\'\i}tez-Llambay}, \& {Beaug{\'e}}}]{ramos.etal.2017}
{Ramos}, X.~S., {Charalambous}, C., {Ben{\'\i}tez-Llambay}, P., \&
  {Beaug{\'e}}, C. 2017, \aap, 602, A101

\bibitem[{{Rein}(2012)}]{rein.2012}
{Rein}, H. 2012, \mnras, 427, L21

\bibitem[{{Rodr{\'\i}guez} {et~al.}(2011){Rodr{\'\i}guez}, {Ferraz-Mello},
  {Michtchenko}, {Beaug{\'e}}, \& {Miloni}}]{rodriguez.etal.2011}
{Rodr{\'\i}guez}, A., {Ferraz-Mello}, S., {Michtchenko}, T.~A., {Beaug{\'e}},
  C., \& {Miloni}, O. 2011, \mnras, 415, 2349

\bibitem[{{Rolf} {et~al.}(2018){Rolf}, {Steinberger}, {Sruthi}, \&
  {Werner}}]{rolf.etal.2018}
{Rolf}, T., {Steinberger}, B., {Sruthi}, U., \& {Werner}, S.~C. 2018, \icarus,
  313, 107

\bibitem[{{Rowe} {et~al.}(2014){Rowe}, {Bryson}, {Marcy}, {Lissauer},
  {Jontof-Hutter}, {Mullally}, {Gilliland}, {Issacson}, {Ford}, {Howell},
  {Borucki}, {Haas}, {Huber}, {Steffen}, {Thompson}, {Quintana}, {Barclay},
  {Still}, {Fortney}, {Gautier}, {Hunter}, {Caldwell}, {Ciardi}, {Devore},
  {Cochran}, {Jenkins}, {Agol}, {Carter}, \& {Geary}}]{rowe.etal.2014}
{Rowe}, J.~F., {Bryson}, S.~T., {Marcy}, G.~W., {et~al.} 2014, \apj, 784, 45

\bibitem[{{Shallue} \& {Vanderburg}(2018)}]{shallue.vanderburg.2018}
{Shallue}, C.~J. \& {Vanderburg}, A. 2018, \aj, 155, 94

\bibitem[{{Shi} {et~al.}(2025){Shi}, {Yang}, {Abbot}, {Liu}, {Kang}, \&
  {Lin}}]{shi.etal.2025}
{Shi}, J., {Yang}, J., {Abbot}, D.~S., {et~al.} 2025, \apj, 989, 139

\bibitem[{{Siegel} \& {Fabrycky}(2021)}]{siegel.fabrycky.2021}
{Siegel}, J.~C. \& {Fabrycky}, D. 2021, \aj, 161, 290

\bibitem[{{Steffen} {et~al.}(2013){Steffen}, {Fabrycky}, {Agol}, {Ford},
  {Morehead}, {Cochran}, {Lissauer}, {Adams}, {Borucki}, {Bryson}, {Caldwell},
  {Dupree}, {Jenkins}, {Robertson}, {Rowe}, {Seader}, {Thompson}, \&
  {Twicken}}]{steffen.etal.2013}
{Steffen}, J.~H., {Fabrycky}, D.~C., {Agol}, E., {et~al.} 2013, \mnras, 428,
  1077

\bibitem[{{Stock} {et~al.}(2020){Stock}, {Nagel}, {Kemmer}, {Passegger},
  {Reffert}, {Quirrenbach}, {Caballero}, {Czesla}, {B{\'e}jar}, {Cardona},
  {D{\'\i}ez-Alonso}, {Herrero}, {Lalitha}, {Schlecker}, {Tal-Or},
  {Rodr{\'\i}guez}, {Rodr{\'\i}guez-L{\'o}pez}, {Ribas}, {Reiners}, {Amado},
  {Bauer}, {Bluhm}, {Cort{\'e}s-Contreras}, {Gonz{\'a}lez-Cuesta}, {Dreizler},
  {Hatzes}, {Henning}, {Jeffers}, {Kaminski}, {K{\"u}rster}, {Lafarga},
  {L{\'o}pez-Gonz{\'a}lez}, {Montes}, {Morales}, {Pedraz}, {Sch{\"o}fer},
  {Schweitzer}, {Trifonov}, {Zapatero Osorio}, \&
  {Zechmeister}}]{stock.etal.2020}
{Stock}, S., {Nagel}, E., {Kemmer}, J., {et~al.} 2020, \aap, 643, A112

\bibitem[{{Su} \& {Lai}(2022)}]{su.lai.2022}
{Su}, Y. \& {Lai}, D. 2022, in AAS/Division of Dynamical Astronomy Meeting,
  Vol.~54, AAS/Division of Dynamical Astronomy Meeting, 302.05

\bibitem[{{Terquem} \& {Papaloizou}(2007)}]{terquem.papaloizou.2007}
{Terquem}, C. \& {Papaloizou}, J. C.~B. 2007, \apj, 654, 1110

\bibitem[{{Williams} \& {Boggs}(2015)}]{williams.boggs.2015}
{Williams}, J.~G. \& {Boggs}, D.~H. 2015, Journal of Geophysical Research
  (Planets), 120, 689

\bibitem[{{Xiang-Gruess} \& {Papaloizou}(2015)}]{xiang-gruess.papaloizou.2015}
{Xiang-Gruess}, M. \& {Papaloizou}, J.~C.~B. 2015, \mnras, 449, 3043

\bibitem[{{Yang} \& {Wei}(2022)}]{yang.wei.2022}
{Yang}, F. \& {Wei}, X. 2022, \pasp, 134, 024401

\end{thebibliography}
\label{LastPage}

\begin{appendix}

\section{Numerical tests of stellar vs. planetary tides}\label{sec:appendix_stellar_vs_planetary}

We put the results of Section \ref{sec:stellar_vs_planetary} to the test simulating the tidal evolution of two different systems, labeled A and B.
System A is a mock system consisting of a 1 $M_\odot$ and 1 $R_\odot$ star, and 3 planets with $m=(1,2,3)\ M_\oplus$, and radii corresponding to an Earth-like density of 5 $gr\,cm^{-3}$.
System B is a truncated replica of K2-138, with its star of 0.93 $M_\odot$ and 0.86 $R_\odot$, and just its three innermost planets, with estimated masses of $m=(3.1,6.3,7.9)\ M_\oplus$ and radii of $R=(1.51,2.30,2.39)\ R_\oplus$ \citep{lopez.etal.2019}.
For both systems, we simulated a simple capture in the 3/2-3/2 intersection, followed with a long-scale tidal evolution.
We show the results of the tidal phase.
We considered both planetary and stellar tides, with stellar $Q'$ values of $10^6$, and $10^2$ for the planets.
This means half the dissipation as that estimated for K2-138 b, but it will be enough for this exercise, the results of which intend to be general.

\begin{figure}[ht!]
\centering
\includegraphics[width=.8\linewidth]{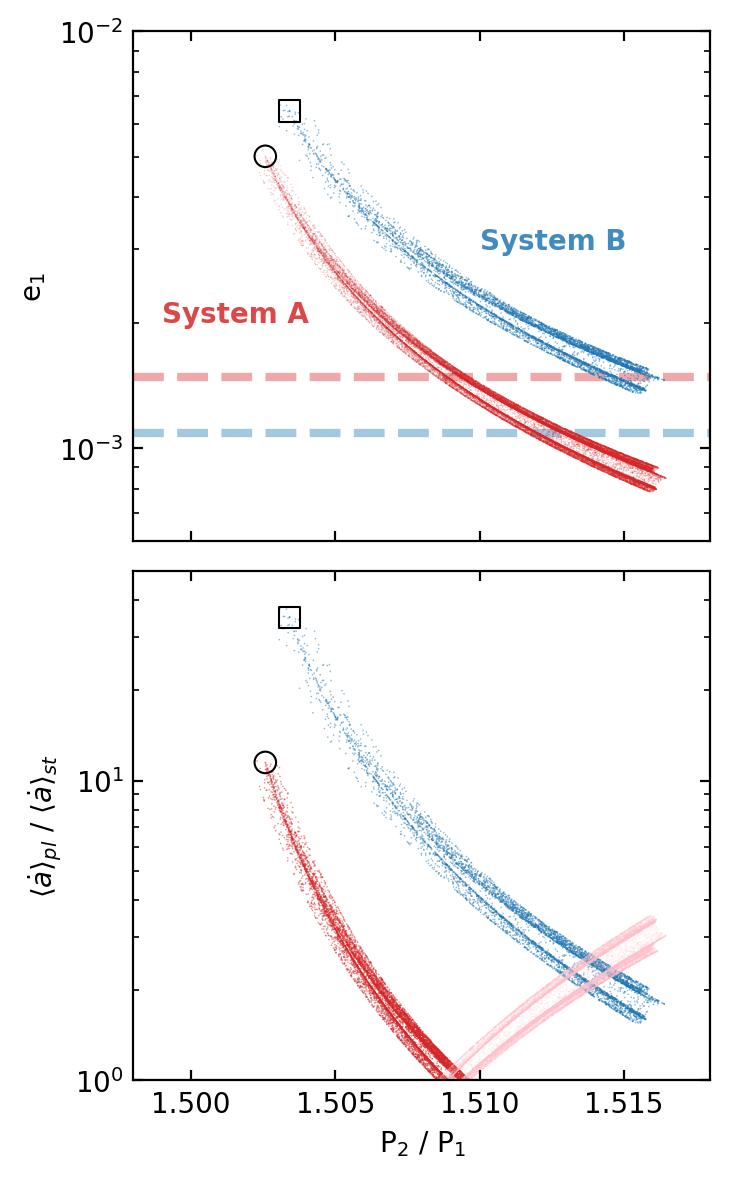}
  \caption{Results of a simulation of the tidal evolution of two resonant chains, labeled Systems A and B (see text for details), plotted in red and blue, respectively.
  The empty shapes mark the start of the tidal evolution. 
  The top panel shows the eccentricity of the innermost planets as a function of the evolving mean-motion ratio $n_1/n_2$.
  Dashed lines mark the critical eccentricity $e_c$ from Equation \eqref{eq:e_1c}.
  The bottom panel shows the analytical ratio of orbital decay contributions stemming from planetary tides and stellar tides, as per Equation \eqref{eq:tau_a.rat}, where we used the $e_1$ values obtained in the simulations (top panel).
  After the red curve crosses the value of one, we flip the ratio and plot the stellar contribution over the planetary one, and paint the curve pink.
  This way, it is easier to gauge that the stellar tides end up being $\sim 3$ times greater.}
  \label{fig:st_vs_pl___n1n2}
\end{figure}

We show the results of the tidal evolution phase in Figure \ref{fig:st_vs_pl___n1n2}, where red curves correspond to System A, and blue curves to System B.
We integrated until the separation of the innermost pairs reached $n_1/n_2\approx1.515$, similar to the observed separation of K2-138's inner pair.
The top panel shows how the eccentricities of the innermost planets evolve during this process.
These start at $5\times10^{-3}$ and $7\times10^{-3}$ (the equilibrium values during the disk phase), and slowly decrease to $8\times10^{-4}$ and $1\times10^{-2}$.
Dashed lines mark the critical eccentricities $e_{1,c}$ below which stellar tides dominate the orbital decay (Equation \ref{eq:e_1c}).
System B has both a higher initial eccentricity and also a lower critical eccentricity.
The former is because of the larger $m_2/m_0$ ratio (see Equation \ref{eq:e1_eq}), while the latter is due to the larger $R_1/R_0$ ratio.
So a bigger second planet, and a smaller star, will keep the eccentricities far above the value $e_c$, and planetary tides will dominate.

In the bottom panel of Figure \ref{fig:st_vs_pl___n1n2}, we used $e_1(t)$ from both simulations to calculate the contribution ratio from Equation (\ref{eq:tau_a.rat}).
In both cases, the contributions from planetary tides to the orbital decay process start being an order of magnitude greater than that of stellar tides.
For system B, this dominance is maintained for the entire simulation, although the ratio does decrease to a value of only $2\sim3$.
System A does eventually cross to the threshold where stellar tides dominate, corresponding to the point where the $e_1$ goes below the critical value (top panel).
For the rest of the curve we plot in pink the inverse of the ratio, that is, the magnitude of the stellar contribution over the planetary one, showing that stellar tides end up being $\sim$ 3 times greater.

These plots show that planetary tides dominate the tidal evolution of resonant chains, especially at the start of the process when eccentricities are higher.
For longer chains, the eccentricity of the first planet as a function of $n_1/n_2$ would look identical, as this is only produced by the 2P-MMR with the second planet.
Nevertheless, the separation process would be slower, so it would spend more of its lifetime close to the initial value, where planetary tides are dominant.
Therefore, long resonant chains can be mainly associated with planetary tides, although markedly small values of $m_2$ and/or high values of $R_0$ can shift the relative importance in favor of stellar tides.

\section{Analytical model for the tidal separation of a resonant chain}
\label{sec:appendix_method}

\cite{papaloizou.2015} developed an analytical model for the tidal evolution of the separation of a resonant chain.
That model is valid for three-planet systems composed of first-order 3P-MMRs.
In this section, we briefly summarize their method, with the slight modification of adding matrices to represent and solve the system of equations.
We will make use of a numerical method to obtain an inverse matrix, but then the solution is analytical.
We will also extend this method for any value of $N$.

Let us consider an $N$-planet system linked by successive first-order 2P-MMRs between adjacent planets, described by the period ratios,

\begin{align}
    \frac{n_i}{n_j}\approx\frac{q_{ij}+1}{q_{ij}}
\end{align},

\noindent where $i=1,...,N$ and $j=i+1$.

We write the Hamiltonian in a Jacobi frame of reference, average over the short-period angles and apply the appropriate transformations from canonical variables to orbital elements.
This yields

\begin{align}\label{eq:H}
    H = H_0 + H_{12}+ ... +H_{N-1,N} = H_0 - R.
\end{align}

Here, $H_0$ is the unperturbed Hamiltonian and $R = -\,(H_{12}\,+\,...\,+\,H_{N-1,N})$ is the perturbation.
The terms $H_{ij}$ are the two-planet resonant contributions expressed as 

\begin{align}
    H_{ij} &= \frac{Gm_j}{a_j}\Big[X^*_{ij} - X^*_{ji}\Big]\frac{1}{\Delta_{ij}^2},\\
    & = n_i^2a_i^2\frac{a_i}{a_j}\frac{m_j}{m_0}\Big[X^*_{ij} - X^*_{ji}\Big]\frac{1}{\Delta_{ij}^2},\notag\\
    & = n_j^2a_j^2\frac{m_j}{m_0}\Big[X^*_{ij} - X^*_{ji}\Big]\frac{1}{\Delta_{ij}^2},\notag
    \smash{\raisebox{0.5\height + 5.7ex}{%
        \hspace{3em}\(\text{$\; j=i+1$}\).
    }}
\end{align}

\noindent where we used $m_0>>m_i$ for $i\in (1,\,...,\,N)$.
We also used $Gm_0\simeq n_i^2a_i^3\simeq n_j^2a_j^3$ to obtain alternate expressions for $H_{ij}$.
We will prefer one or the other depending on the equation.
We introduced the variables $X^*_{ij} = \Delta_{ij}^2 D_{ij} e_i \cos(\phi_{ij})$.
In addition, we define their sine counterparts as $X_{ij} = \Delta_{ij}^2 D_{ij} e_i \sin(\phi_{ij})$.
$X_{ji}$ and $X^*_{ji}$ are analogous, after replacing the sub indexes.
The $\Delta_{ij}$ quantities are a measure of the distance from resonance, and they are defined as $\Delta_{ij}=\Delta_{ji}=-q_{ij}n_i + (q_{ij}+1)n_j$.
This quantity is 0 at exact commensurability, and grows to negative values as the system undergoes tidal separation.
The resonant angles are given by $\phi_{ij}=\theta_{ij}+\varpi_i$, where
$\theta_{ij}=\theta_{ji}=q_{ij}\lambda_i - (q_{ij}+1)\lambda_j$.
Finally, the $D_{ij}$ coefficients are given by

\begin{flalign}
  &D_{ij}=\frac12\Bigg(\alpha_{ij}\frac{db^{q_{ij}+1}_{1/2}(\alpha)}{d\alpha}\Bigg\rvert_{\alpha_{ij}}+\quad 2(q_{ij}+1)b^{q_{ij}+1}_{1/2}(\alpha_{ij})\,\Bigg),&\\
  &D_{ji} = \frac12
    \Bigg(\alpha_{ij}\,\frac{d b^{q_{ij}}_{1/2}(\alpha)}{d\alpha}\Bigg\rvert_{\alpha_{ij}}
      +\quad (2q_{ij}+1)\,b^{q_{ij}}_{1/2}(\alpha_{ij}),\\
    &\qquad\qquad\qquad\qquad\qquad\ -\ \ \ \ (2q_{ij}+2)\,\alpha_{ij}\,\delta_{q_{ij},1}\,\Bigg)\notag,&
\end{flalign}

\noindent where the $b_s^j(\alpha)$ factors are the usual Laplace coefficients \citep{brouwer.clemence.1961}, $\alpha_{ij} = a_i/a_j$, and $\delta_{i,j}$ is a Dirac delta function.

We take Lagrange's planetary equations for $\dot e_i,\ \dot n_i$, and $\dot \phi_{ij}$ to first order in eccentricities.
We also add a tidal circularization term to $\dot{e}_i$, and its corresponding effect on $\dot{n}_i$, as consequence of the conservation of angular momentum,

\begin{align}
    \dv{e_i}{t} &= -\frac{1}{n_ia_i^2e_i}\pdv{R}{\varpi_i} -\frac{e_i}{\tau_{e,i}}\label{eq:lagrange_e},\\
    \dv{n_i}{t} &= -\frac{3}{a_i^2}\pdv{R}{\lambda_i} +\frac{3n_i e_i^2}{\tau_{e,i}}\label{eq:lagrange_n},\\
    \dv{\phi_{ij}}{t} &\approx \Delta_{ij} - \frac{1}{n_ia_i^2e_i}\pdv{R}{e_i}\label{eq:lagrange_phi}.
\end{align}

For the following elaboration, we consider $N=4$.
We do this for the sake of clarity, as the description of the matrices can become quite abstract and complicated.
Nevertheless, all equations and matrices are repetitive and formulaic, so their extension from $N=4$ is trivial.

We also consider that the planets are linked via 3P-MMRs, which also restricts the mean-motions.
For the case of $N=4$, there are $N-2=2$ relations, given by

\begin{align}
    q_{12}n_1-(q_{12}+q_{23}+1)n_2+(q_{23}+1)n_3 = 0\label{eq:3pmmr_4pl_123},\\
    q_{23}n_2-(q_{23}+q_{34}+1)n_3+(q_{34}+1)n_4 = 0\label{eq:3pmmr_4pl_234},
\end{align}

\noindent which, after some rearranging, imply that $\Delta_{12}=\Delta_{23}=\Delta_{34}$.
Therefore, we can replace all of the previous $\Delta_{ij}$ by $\Delta_{12}$, which is the variable we ultimately want to solve for.

We start by calculating the equilibrium eccentricities which stem from the 2P-MMRs.
These will allow us to replace eccentricities by a combination of fixed parameters and also the total separation $\Delta_{12}$.
We will consider the equations of motion for the resonant angles (Eq. \ref{eq:lagrange_phi}) expressed as

\begin{flalign}\label{eq:dphi12dt_4pl}
    &\dv{\phi_{12}}{t} = \Delta_{12}
    + \frac{n_1}{e_1^2\Delta_{12}^2}\bigg(\frac{a_1}{a_2}\frac{m_2}{m_0}X^*_{12}\bigg),&
\end{flalign}

\begin{flalign}\label{eq:dphi21dt_4pl}
    &\dv{\phi_{21}}{t} = \dv{\phi_{23}}{t} = \Delta_{12} + \frac{n_2}{e_2^2\Delta_{12}^2}
    \bigg(-\frac{m_1}{m_0}X^*_{21} + \frac{a_2}{a_3}\frac{m_3}{m_0}X^*_{23}\bigg),&
\end{flalign}

\begin{flalign}\label{eq:dphi32dt_4pl}
    &\dv{\phi_{32}}{t} = \dv{\phi_{34}}{t} = \Delta_{12} + \frac{n_3}{e_3^2\Delta_{12}^2}
    \bigg(-\frac{m_2}{m_0}X^*_{32}+ \frac{a_3}{a_4}\frac{m_4}{m_0}X^*_{34}\bigg),&
\end{flalign}

\begin{flalign}\label{eq:dphi43dt_4pl}
    &\dv{\phi_{43}}{t} = \Delta_{12}
    + \frac{n_4}{e_4^2\Delta_{12}^2}\bigg(-\frac{m_3}{m_0}X^*_{43}\bigg).&
\end{flalign}

We then consider that, as the triplet moves away from the intersection of 2P-MMRs, these resonances still hold, so the above derivatives are zero.
We also assume the angles will have reached symmetric equilibrium values of 0 (for $\phi_{12}$, $\phi_{23}$ and $\phi_{34}$) and $\pi$ (for $\phi_{21}$, $\phi_{32}$ and $\phi_{43}$).
Then, $X^*_{ij} = \Delta_{12}^2 D_{ij} e_i$ and $X^*_{ji} = -\Delta_{12}^2 D_{ji} e_j$.
The selected symmetric resonant centers are required to keep the eccentricity positive, given that $\Delta_{12}$ is expected to be negative as period ratios increase.
Nevertheless, the alternate case where the centers are switched would produce a change of signs which would be irrelevant, given that our eccentricities enter the model as squared, as we show below. 
Therefore, we can solve for the equilibrium eccentricities, which yield

\begin{flalign}
    &e_1=-\frac{1}{\Delta_{12}}\bigg[n_1\Big(\frac{m_2}{m_0}\frac{a_1}{a_2}D_{12}\Big)\bigg] = -\frac{1}{\Delta_{12}}E_1\label{eq:e1_eq},&\\
    &e_2=-\frac{1}{\Delta_{12}}\bigg[n_2\Big(\frac{m_3}{m_0}\frac{a_2}{a_3}D_{23}+\frac{m_1}{m_0}D_{21}\Big)\bigg] = -\frac{1}{\Delta_{12}}E_2\label{eq:e2_eq},&\\
    &e_3=-\frac{1}{\Delta_{12}}\bigg[n_3\Big(\frac{m_4}{m_0}\frac{a_3}{a_4}D_{34} + \frac{m_2}{m_0}D_{32}\Big)\bigg] = -\frac{1}{\Delta_{12}}E_3\label{eq:e3_eq},&\\
    &e_4=-\frac{1}{\Delta_{12}}\bigg[n_4\Big(\frac{m_3}{m_0}D_{43}\Big)\bigg] = -\frac{1}{\Delta_{12}}E_4.&\label{eq:e4_eq}
\end{flalign}

As we wish to solve for $\Delta_{12}= -q_{12}n_1 + (q_{12}+1)n_2$, we follow \cite{papaloizou.2015} and calculate $\Delta_{12}^2\dot{\Delta}_{12}$, given by

\begin{align}\label{eq:Delta2_dDeltadt_pre}
\Delta_{12}^2\dv{\Delta_{12}}{t} = -q_{12}\Delta_{12}^2\dv{n_1}{t}+(q_{12}+1)\Delta_{12}^2\dv{n_2}{t}
\end{align}

\noindent for which we use the equations of motion for $\dot{n}_i$ of Eq. \eqref{eq:lagrange_n}

\begin{flalign}\label{eq:dn1dt}
    &\dv{n_1}{t} =\ \ \,\nu_{12}\Big(X_{12}-X_{21}\Big)\frac{1}{\Delta_{12}^2} + \frac{3n_1e_1^2}{\tau_{e,1}}&
\end{flalign}

\begin{flalign}\label{eq:dn2dt}
    &\dv{n_2}{t} = -\nu_{21}\Big(X_{12}-X_{21}\Big)\frac{1}{\Delta_{12}^2} &\\
    &\qquad\ \ + \nu_{23}\Big(X_{23}-X_{32}\Big)\frac{1}{\Delta_{23}^2} + \frac{3n_2e_2^2}{\tau_{e,2}}&\notag
\end{flalign}

\begin{flalign}\label{eq:dn3dt}
    &\dv{n_3}{t} = -\nu_{32}\Big(X_{23}-X_{32}\Big)\frac{1}{\Delta_{23}^2} &\\
    &\qquad\ \ + \nu_{34}\Big(X_{34}-X_{43}\Big)\frac{1}{\Delta_{34}^2} + \frac{3n_3e_3^2}{\tau_{e,3}}&
\end{flalign}

\begin{flalign}\label{eq:dn4dt}
    &\dv{n_4}{t} = -\nu_{43}\Big(X_{34}-X_{43}\Big)\frac{1}{\Delta_{34}^2} + \frac{3n_4e_4^2}{\tau_{e,4}}&
\end{flalign}

\noindent where we defined

\begin{align}
    \nu_{ij} &= 3q_{ij}n_i^2\frac{m_j}{m_0}\frac{a_i}{a_j}\label{eq:nu_12}\\
    \nu_{ji} &= 3(q_{ij}+1)n_j^2\frac{m_i}{m_0}\label{eq:nu_21}
    \smash{\raisebox{0.5\height + 1.5ex}{\hspace{3em}\(\text{$\quad j=i+1$}\)}}
\end{align}

We introduce the previous derivatives in Equation \eqref{eq:Delta2_dDeltadt_pre}

\begin{flalign}\label{eq:Delta2_dDeltadt}
    &\Delta_{12}^2\dv{\Delta_{12}}{t} 
    =-\ \Big(X_{12}-X_{21}\Big)\,\Big(q_{12}\nu_{12}+(q_{12}+1)\nu_{21}\Big) &\\ 
    &\qquad\qquad\ \,\, + \Big(X_{23}-X_{32}\Big)\,(q_{12}+1)\nu_{23} \notag &\\
    &\qquad\qquad\ \,\, - \frac{3q_{12}n_1 E_1^2}{\tau_{e,1}} + \frac{3(q_{12}+1)n_2 E_2^2}{\tau_{e,2}} ,\notag &
\end{flalign}

To solve for $\Delta_{12}(t)$, we need to solve for all of the $2\,(N-1)\ $ $X_{ij}$ variables; in this case, we need $2\,(N-1)=6$ equations. For that total, $N$ equations will be taken from the individual eccentricities (Eq. \ref{eq:lagrange_e})

\begin{flalign}\label{eq:de1dt}
    &\dv{e_1}{t} = n_1\frac{a_1}{a_2}\frac{m_2}{m_0}\frac{X_{12}}{\Delta_{12}^2 e_1} - \frac{e_1}{\tau_{e,1}},&
\end{flalign}

\begin{flalign}\label{eq:de2dt}
    &\dv{e_2}{t} = - n_2\frac{m_1}{m_0}\frac{X_{21}}{\Delta_{12}^2 e_2} + n_2\frac{a_2}{a_3}\frac{m_3}{m_0}\frac{X_{23}}{\Delta_{23}^2 e_2} - \frac{e_2}{\tau_{e,2}},&
\end{flalign}

\begin{flalign}\label{eq:de3dt}
  &\dv{e_3}{t} = - n_3\frac{m_2}{m_0}\frac{X_{32}}{\Delta_{23}^2 e_3} + n_3\frac{a_3}{a_4}\frac{m_4}{m_0}\frac{X_{34}}{\Delta_{12}^2 e_3} - \frac{e_3}{\tau_{e,3}},&
\end{flalign}

\begin{flalign}\label{eq:de4dt_4pl}
  &\dv{e_4}{t} = - n_4\frac{m_3}{m_0}\frac{X_{43}}{\Delta_{12}^2 e_4} - \frac{e_4}{\tau_{e,4}}.&
\end{flalign}

We are interested in the secular evolution of the system, by which these derivatives would be close to 0.
We then multiply by $\Delta_{12}^2 e_i$ to get

\begin{flalign}
    E_1^2&=\gamma_{12}X_{12}\label{eq:e1_2} &= \Bigg(E_1^2\tau_{e,1}\Bigg)\ \tau_{e,1}^{-1},& &\\
    E_2^2&=-\gamma_{21}X_{21} + \gamma_{23}X_{23} &= \Bigg(E_2^2\tau_{e,2}\Bigg)\ \tau_{e,2}^{-1}\label{eq:e2_2},& &\\
    E_3^2&=-\gamma_{32}X_{32} + \gamma_{34}X_{34}&= \Bigg(E_3^2\tau_{e,3}\Bigg)\ \tau_{e,3}^{-1}\label{eq:e3_2},& &\\
    E_4^2&=-\gamma_{43}X_{43} &= \Bigg(E_4^2\tau_{e,4}\Bigg)\ \tau_{e,4}^{-1}\label{eq:e4_2}.& &
\end{flalign}

\noindent where we defined
\begin{align}
    &\gamma_{ij} = \tau_{e,i}n_i\frac{m_j}{m_0}\frac{a_i}{a_j}
      \label{eq:gamma_12},\\[0pt]
    &\gamma_{ji} = \tau_{e,j}n_j\frac{m_i}{m_0}
      \label{eq:gamma_21}
    \smash{\raisebox{0.5\height + 1.5ex}{\hspace{3em}\(\text{$\quad j=i+1$}\)}},
\end{align}

The remaining two equations will be assembled by taking the time derivatives of the 3P-MMR relations in Eqs. \eqref{eq:3pmmr_4pl_123} and \eqref{eq:3pmmr_4pl_234}, and introducing Eqs. \eqref{eq:dn1dt}-\eqref{eq:dn4dt}.
Then, we get

\begin{align}\label{eq:deriv_3p-mmr}
    &\beta_{inn}^{(123)}X_{12}\ -\beta_{inn}^{(123)}X_{21} + \beta_{out}^{(123)}X_{23}\ -\beta_{out}^{(123)}X_{32}, &\\
    &=\Bigg(q_{12}n_1 E_1^2\Bigg)\ \tau_{e,1}^{-1} - \Bigg((q_{12}+q_{23}+1)n_2 E_2^2\Bigg)\ \tau_{e,2}^{-1}&\notag\\
    &\quad\Bigg((q_{23}+1)n_3 E_3^2\Bigg)\ \tau_{e,3}^{-1},&\notag
\end{align}

\begin{align}\label{eq:deriv_3p-mmrb}
    &\beta_{inn}^{(234)}X_{23}\ -\beta_{inn}^{(234)}X_{32} + \beta_{out}^{(234)}X_{34}\ -\beta_{out}^{(234)}X_{43} &\\
    &=\Bigg(q_{23}n_2 E_2^2\Bigg)\ \tau_{e,2}^{-1} - \Bigg((q_{23}+q_{34}+1)n_3 E_3^2\Bigg)\ \tau_{e,3}^{-1},&\notag\\
    &\quad\Bigg((q_{34}+1)n_4 E_4^2\Bigg)\ \tau_{e,4}^{-1},&\notag
\end{align}

\noindent where we define

\begin{align}
    \beta_{inn}^{(ijk)} &= -q_{ij}\nu_{ij} + (q_{ij}+q_{jk}+1)\nu_{ji}\label{eq:beta_inn},\\
    \beta_{out}^{(ijk)} &= (q_{ij}+q_{jk}+1)\nu_{jk} - (q_{jk}+1)\nu_{kj}\label{eq:beta_out},
\end{align}

\noindent where $j=i+1$ and $k=i+2$.

We can then assemble this system of equations in matrix form,

\begin{flalign}
    \invisibleunderbrace{\mathbf{A}}{2\,(N-1)\,\times\,2\,(N-1)}
    \begin{pmatrix}
        X_{12}\\
        X_{21}\\
        X_{23}\\
        X_{32}\\
        X_{34}\\
        X_{43}
    \end{pmatrix}
    =\invisibleunderbrace{\mathbf{B}}{2\,(N-1)\,\times\,N}
    \begin{pmatrix}
        \tau_{e,1}^{-1} \\[0.2cm]
        \tau_{e,2}^{-1} \\[0.2cm]
        \tau_{e,3}^{-1} \\[0.2cm]
        \tau_{e,4}^{-1} \\
    \end{pmatrix},
\end{flalign}

\noindent where the expressions for matrices $\mathbf{A}$ and $\mathbf{B}$ are given in Eq. \eqref{eq:matrices_def}.
Each matrix is composed of two vertically stacked sub-matrices which are formulaic and easily extendable.
Extending matrix $\mathbf{A}$ requires extending the slanted diagonal with a pair of ($-\gamma_{ij}$,$\gamma_{ji}$) for each row (except at both ends, where the pairs are truncated), and adding additional rows at the bottom with 4 $\beta^{ijk}$ values per triplet.
Extending matrix $\mathbf{B}$ is even easier, as it has a diagonal matrix on top, and the one below is just the 3P-MMRs relations multiplied by the $E_i^2$ factors.

Next, we solve for the $X_{ij}$ variables.
We multiply by $\mathbf{A^{-1}}$ and produce $\mathbf{M} = \mathbf{A^{-1}}\mathbf{B}$, of shape $2(N-1)\times\, N$.
The process of finding $\mathbf{A^{-1}}$ is done numerically via the Gauss-Jordan method.
Finally, we have

\begin{figure*}[!t]
  \centering
  \begin{minipage}{\textwidth}
    \begin{equation}\label{eq:matrices_def}
      \begin{aligned}
        \invisibleunderbrace{\mathbf{A}}{2(N-1)\times 2(N-1)}
        &=
        \begin{array}{@{}c@{\;}c@{}}
          \begin{array}{@{}r@{\;}c@{}}
            N & \left\{ \vphantom{\begin{array}{c}
              \g[12]\\[0.15cm]\g[21]\\[0.15cm]\g[32]\\[0.15cm]\g[43]
            \end{array}}\right.\\[1.5ex]
            N-2 & \left\{ \vphantom{\begin{array}{c}
              \betainn{123}\\[0.15cm]\betainn{234}
            \end{array}}\right.
          \end{array}
          &
          \overbrace{
            \begin{pmatrix}
              \g[12] & 0 & 0 & 0 & 0 & 0 \\[0.15cm]
              0 & -\g[21] & \g[23] & 0 & 0 & 0 \\[0.15cm]
              0 & 0 & 0 & -\g[32] & \g[34] & 0 \\[0.15cm]
              0 & 0 & 0 & 0 & 0 & -\g[43] \\[0.15cm]
              \betainn{123} & -\betainn{123} & \betaout{123} & -\betaout{123} & 0 & 0 \\[0.15cm]
              0 & 0 & \betainn{234} & -\betainn{234} & \betaout{234} & -\betaout{234}
            \end{pmatrix},
          }^{\displaystyle 2(N-1)}
        \end{array}
        \\[1.2cm] 
        \invisibleunderbrace{\mathbf{B}}{2(N-1)\times N}
        &=
        \begin{array}{@{}c@{\;}c@{}}
          \begin{array}{@{}r@{\;}c@{}}
            N & \left\{ \vphantom{\begin{array}{c}
              1\\[0.15cm]1\\[0.15cm]1\\[0.15cm]1
            \end{array}}\right.\\[1.5ex]
            N-2 & \left\{ \vphantom{\begin{array}{c}
              \frac{q_{12}n_1}{\tau_{e,1}}\\[0.15cm]\frac{q_{12}n_1}{\tau_{e,1}}
            \end{array}}\right.
          \end{array},
          &
          \overbrace{
            \begin{pmatrix}
              \tau_{e,1}E_1^2 & 0 & 0 & 0 \\[0.15cm]
              0 & \tau_{e,2}E_2^2 & 0 & 0 \\[0.15cm]
              0 & 0 & \tau_{e,3}E_3^2 & 0 \\[0.15cm]
              0 & 0 & 0 & \tau_{e,4}E_4^2 \\[0.15cm]
              q_{12}n_1E_1^2 &
              -(q_{12}+q_{23}+1)n_2E_2^2 &
              (q_{23}+1)n_3E_3^2 &
              0\\[0.15cm]
              0 &
              q_{23}n_2E_2^2 &
              -(q_{23}+q_{34}+1)n_3E_3^2 &
              (q_{34}+1)n_4E_4^2
            \end{pmatrix}
          }^{\displaystyle N}
        \end{array},
      \end{aligned}
    \end{equation}
  \end{minipage}
\end{figure*}

\begin{align}
    \begin{pmatrix}
        X_{12}\\
        X_{21}\\
        X_{23}\\
        X_{32}\\
        X_{34}\\
        X_{43}
    \end{pmatrix}
    =\mathbf{A^{-1}B}
    \begin{pmatrix}
        \tau_{e,1}^{-1} \\[0.2cm]
        \tau_{e,2}^{-1} \\[0.2cm]
        \tau_{e,3}^{-1} \\[0.2cm]
        \tau_{e,4}^{-1} \\
    \end{pmatrix}
    =\invisibleunderbrace{\mathbf{M}}{2(N-1)\,\times\,N}
    \begin{pmatrix}
        \tau_{e,1}^{-1} \\[0.2cm]
        \tau_{e,2}^{-1} \\[0.2cm]
        \tau_{e,3}^{-1} \\[0.2cm]
        \tau_{e,4}^{-1} \\
    \end{pmatrix}.
  \label{eq:Xij_matrix}
\end{align}

Now we can introduce $X_{12}$, $X_{21}$, $X_{23}$, and $X_{34}$ into Eq. \eqref{eq:Delta2_dDeltadt}.
We note that regardless of $N$, only these four $X_{ij}$ variables are required for this calculation.
We can then calculate $\Delta_{12}^2\ d\Delta_{12}/dt = 1/3\ d(\Delta_{12}^3)/dt$.
This quantity will have units of mean-motion cubed over time, so we follow \cite{papaloizou.2015}\footnote{Note we use $n_2$ instead of $n_1$.
This is because $\Delta_{12}/n_2$ is linear in $n_1/n_2$, giving less convoluted solutions.} and propose a simple solution accordingly, introducing a constant timescale, $T$, via

\begin{align}
    \Delta_{12}^2\dv{\Delta_{12}}{t} = \frac{1}{3}\dv{}{t}\left(\Delta_{12}^3\right) = -\frac{n_2^3}{3T} \ \Bigg(\frac{q_{12}+1}{100}\Bigg)^3
    \label{eq:ddt(Delta_12^3)},
\end{align}

\noindent where the last parenthesis was added to ensure that $T$ would represent the time it would take for the resonant chain to increase its $n_1/n_2$ separation in a magnitude of 1$\%$ starting from exact commensurability (i.e increase 1.5 to 1.515, for a 3/2 2P-MMR).
These are ostensibly large separations but comparable to those in real chains such as K2-138 ($n_1/n_2\approx 1.513$), and generally places $T$ in the manageable scale of a few Gyr.
We calculate $T$ by solving it from the previous equation,

\begin{align}
  T = -\frac{n_2^3}{3(\Delta_{12}^2\ \dot{\Delta}_{12})}\ \bigg(\frac{q_{12}+1}{100}\bigg)^3
  \label{eq:T},
\end{align}

\noindent where the left denominator was calculated using the $X_{ij}$ variables and Equation \eqref{eq:Delta2_dDeltadt}.

We can also solve Equation \eqref{eq:T} to get the solution for the time evolution of the mean-motion difference, yielding

\begin{align}
    \Delta_{12}(t) = -n_2\frac{q_{12}+1}{100}\left(\frac{t}{T}\right)^{\frac{1}{3}}
\label{eq:Delta_12_solution},
\end{align}

\noindent for which we assumed that $\Delta_{12}(0) = 0$, meaning that the triplets start from the exact 2P-MMR intersection.
Although large enough initial offsets can affect our results, those remaining from disk interaction (as predicted by current models) are too small compared with those observed in long resonant chains.
In those cases where our estimation of an upper bound for disk offset is imprecise, young enough systems may require consideration of the initial separation.
This may involve a non-negligible portion of the resonant chain population \citep{dai.etal.2024,hamer.schlaufman.2024}.
We elaborate further on this choice in Section \ref{sec:init_offsets}.
If needed, an initial relative offset $f_{12}(0)$ can be introduced to our solutions by simply replacing $t\rightarrow t+t_0$, where

\begin{align}
  t_0 = \big(100\ f_{12}(0)\big)^3\ T
  \label{eq:t0}.
\end{align}

Ultimately, we want to study the separation by the evolution of the mean-motion ratio, whose solution we can solve from the definition of $\Delta_{12}$. Therefore, we have

\begin{align}\label{eq:n1n2_solution_app}
    &\frac{n_1}{n_2}\big(t\big) = \frac{q_{12}+1}{q_{12}}\ \Bigg(1 + \frac{1}{100}\left(\frac{t}{T}\right)^{\frac{1}{3}}\Bigg),&
\end{align}

\noindent where we can see that $n_1/n_2(t=T) = 1.01\ (q_{12}+1)/q_{12}$, as explained earlier.

In summary, in this method we use Equation \eqref{eq:Xij_matrix} to calculate the $X_{ij}$ variables, then use those to calculate $\Delta_{12}^2\dot{\Delta}_{12}$ from Equation \eqref{eq:Delta2_dDeltadt}, and then solve for the timescale $T$ from Equation \eqref{eq:T}.
Finally, we use Equation \eqref{eq:n1n2_solution_app} to calculate the evolution of the separation between planets 1 and 2, which can be extended to the other pairs via the 3P-MMR relations.

\end{appendix}

\end{document}